\newcommand{\be}{\begin{equation}}
\newcommand{\ee}{\end{equation}}
\newcommand{\bea}{\setlength\arraycolsep{2pt} \begin{eqnarray}}
\newcommand{\eea}{\end{eqnarray}}
\newcommand{\nn}{\nonumber}

\def\ft#1#2{{\textstyle{\frac{\scriptstyle #1}{\scriptstyle #2} } }}
\def\fft#1#2{{\frac{#1}{#2}}}

\def\0{{\sst{(0)}}}
\def\1{{\sst{(1)}}}
\def\2{{\sst{(2)}}}
\def\3{{\sst{(3)}}}
\def\4{{\sst{(4)}}}
\def\5{{\sst{(5)}}}
\def\6{{\sst{(6)}}}
\def\7{{\sst{(7)}}}
\def\8{{\sst{(8)}}}
\def\sst#1{{\scriptscriptstyle #1}}

\makeatletter
\@ifundefined{@parse@version@dash}{%
\def\@parse@version#1{\@parse@version@0#1}
\def\@parse@version@#1/#2/#3#4#5\@nil{%
\@parse@version@dash#1-#2-#3#4\@nil}
\def\@parse@version@dash#1-#2-#3#4#5\@nil{%
  \if\relax#2\relax\else#1\fi#2#3#4 }
}{}
\makeatother

\documentclass[%
 reprint,
 amsmath,amssymb,
 aps,
twocolumn,
preprintnumbers
]{revtex4-2}

\usepackage{graphicx}
\usepackage{dcolumn}
\usepackage{bm}
\usepackage{color}
\usepackage{tabularx}
\usepackage{hyperref}

\usepackage[vcentermath]{youngtab}

\def\.{{\cdot}}

\newcommand{\lsim}{\mathrel{\hbox{\rlap{\lower.55ex \hbox{$\sim$}} \kern-.3em \raise.4ex \hbox{$<$}}}}
\newcommand{\gsim}{\mathrel{\hbox{\rlap{\lower.55ex \hbox{$\sim$}} \kern-.3em \raise.4ex \hbox{$>$}}}}

\def\be#1\ee{\begin{align}#1\end{align}}

\renewcommand\hat[1]{\widehat{#1}}

\definecolor{byzantine}{rgb}{0.74, 0.2, 0.64}

\begin{document}

\preprint{CALT-TH 2022-17}

\title{Flat-space structure of gluon and graviton in AdS}


\author{Yue-Zhou Li$^{\mathcal{E},\mathcal{P}}$}
\affiliation{${}^{\mathcal{E}}$Department of Physics, McGill University, 3600 Rue University, Montr\'eal, H3A 2T8, QC Canada \\
${}^{\mathcal{P}}$HKUST Jockey Club Institute for Advanced Study,
The Hong Kong University of Science and Technology, Clear Water Bay, Hong Kong, P.R.China\\
}


\begin{abstract}

We report the differential representation of three-point and four-point amplitudes for Yang-Mills fields and Einstein gravity in AdS at tree-level. The amplitudes exhibit the flat-space structures by using the weight-shifting operators with reordering, which makes the differential double copy relation at the three-point level straightforward. For four-point Yang-Mills amplitudes, we establish the differential BCJ relation, which can be useful for proving the differential doubly copy at the four-point level in the future.

\end{abstract}

\maketitle

\section{Introduction}

Scattering theory plays an essential role in understanding the fundamental principles of particles. For past decades, there has been huge progress in scattering theory in flat-space, which not only successfully predicts and explains many exciting discoveries made by collider \cite{Weinberg:1995mt}, but also remarkably reveals hidden structures linked to the entity of local quantum field theories, such as Bern-Carrasco-Johansson (BCJ) relations \cite{Bern:2008qj} and the double copy structures \cite{KAWAI19861,Bern:2010ue,Bern:2019prr}. 

However, our universe is not flat. Although our universe is generally curved, the local scattering of particles in a small regime compared to the curvature can still be approximated by flat-space physics, as verified by scattering experiments. It is then natural and crucial to ask, do those beautiful structures of scattering remain for curved spacetime? How good can the locally flat scattering experiments say about curved spacetime? A natural starting point to answer these questions would be studying other maximally symmetric spacetimes, such as Anti de-Sitter (AdS) and de-Sitter (dS) space. 

AdS scattering is studied extensively due to its correspondence with conformal field theories (CFT) \cite{Maldacena:1997re}. The unitary AdS physics can be explored by using unitary large-$N$ CFT \footnote{For dS, although it turns out its essential structures are similar to AdS and analytic continuation exists to go from one to another \cite{Sleight:2020obc}, the unitarity of dS scattering is not a standard concept from CFT perspective \cite{Hogervorst:2021uvp}.} that is highly constrained by conformal symmetry and crossing symmetry (see \cite{Hartman:2022zik, Bissi:2022mrs} for quick reviews). As expected, the appropriate limit of conformal correlators and conformal data corresponds to the flat-space limit of local AdS scattering, giving back the flat-space scattering data (see, e.g., \cite{Okuda:2010ym, Penedones:2010ue, Fitzpatrick:2011hu, Maldacena:2015iua, Raju:2012zr, Paulos:2016fap, Komatsu:2020sag, Hijano:2019qmi, Hijano:2019qmi, Li:2021snj} and references therein) \footnote{However, there exist exemptions for special analytic regimes that are not well-understood yet, see \cite{Komatsu:2020sag} for recent explorations.}. However, previously it is usually not expected to reconstruct full AdS amplitudes/CFT correlators from flat-space. In this sense, the flat-space limit of AdS/CFT is crucially different from ``flat holography'' \cite{Bagchi:2016bcd}. Recent progress was made in \cite{Eberhardt:2020ewh, Roehrig:2020kck, Diwakar:2021juk, Herderschee:2022ntr, Cheung:2022pdk,Gomez:2021qfd,Gomez:2021ujt,Herderschee:2021jbi}, which surprisingly found the differential representation of scalar (A)dS amplitudes by writing (A)dS amplitudes as conformal generators acting on scalar contact Witten diagrams. This differential representation not only makes flat-space limit manifest but also allows one to uplift the flat-space amplitudes to (A)dS in a universal way.

In this paper, we aim to progress toward extending the differential representation to spinning correlators by focusing on massless gluon and graviton in AdS from Yang-Mills (YM) theory and Einstein gravity. Massless spinning particles in flat-space are constrained by stringent consistency conditions and encode hidden structures such as double copy \cite{KAWAI19861, Bern:2010ue, Bern:2019prr}. We report the differential representation for YM and graviton amplitudes in AdS. We show we can uplift gluon and graviton amplitudes in AdS from flat-space up to a finite number of additional contact structures. We argue and expect that the additional contact structures can be bootstrapped by requiring the conservation of conserved currents and stress-tensors. The same arguments also apply to hidden structures like BCJ relations and the double copy, which are now in a differential format. Although we start our exploration in AdS, our results should be readily translated to dS \cite{Maldacena:2002vr}.

The rest of this paper is organized as follows. In section \ref{sec: building blocks}, we introduce our differential operators for representing spinning correlators and comment on power counting principles in CFT. In section \ref{sec: diff rep}, we construct the differential representation for three-point and four-point YM amplitudes and graviton amplitudes; we show that the double copy is straightforward for three-point amplitudes. In section \ref{sec: BCJ}, we propose another differential representation for YM amplitudes, which allows us to uplift flat-space BCJ numerators and prove the differential BCJ relations. We summarize and point out future directions in section \ref{sec: summary}. We record detailed ingredients in our derivations in appendix \ref{app: vertex} and \ref{app: ddh rep and id}.

\noindent {\bf Note:} During the preparation of this work, \cite{Lee:2022fgr} appeared, which has partial overlap with the idea of using weight-shifting operators and the discussions on three-point double copy in section \ref{sec: building blocks} and subsection \ref{subsec: three-pt}.

\section{Building blocks for differential representation}
\label{sec: building blocks}

This section introduces our notations and building blocks for differentially representing conserved current and stress-tensor correlators. 

For scalars, it is easy to find the differential representation for any contact diagrams up to any points by using the coset construction \cite{Cheung:2022pdk}. The spirit is that the contracted bulk derivatives can be replaced by the contractions of conformal generators, as guaranteed by the conformal symmetry. For spinning objects, the tensor structures appear, which can either contract among themselves or with bulk derivatives. From the perspective of CFTs, the spinning indices shall be captured by the spin-up weight-shifting operator \cite{Karateev:2017jgd} (written in terms of embedding formalism \cite{Costa:2011mg})
\be
\mathcal{D}^{0+}_{\mu}= (J+\Delta)Z_\mu+X_\mu Z\cdot\partial_X\,,
\ee
where $X, Z$ are coordinates and polarizations in the embedding space of CFTs, they obey $X^2=Z^2=X\cdot Z\equiv 0$. $\Delta$ and $J$ are the scaling dimension and spin of the operators it acts on, On the other hand, we speculate and show that bulk derivatives can be replaced by dimension-up weight-shifting operator $\mathcal{D}^{+0}$ (that raises the scaling dimensions \cite{Karateev:2017jgd}) modulo bulk coordinates
\begin{align}
 \mathcal{D}^{+0}_{\mu}& = c_1 \partial_{X^\mu}+c_2 X_\mu \partial_X^2+c_3 Z_\mu \partial_Z\cdot\partial_X+c_4 Z\cdot\partial_X \partial_{Z^\mu}\nn\\
& 
+c_5 X_\mu Z\cdot \partial_X \partial_Z\cdot\partial_X + c_6 Z_\mu Z\cdot\partial_X \partial_Z^2
\nn\\
& +c_7 X_\mu (Z\cdot\partial_X)^2 \partial_Z^2\,,
\end{align}
where the coefficients can be found in \cite{Karateev:2017jgd}. An intuitive way to convince ourselves that the dimension-up weight-shifting operator plays a role like momentum in flat-space is that the flat-space momentum is $i\partial/\partial x$ which also increases the ``scaling dimensions''. In this paper, we find that it is instructive to define the following differential operators proportional to weight-shifting operators with state-dependent normalizations
\begin{align}
& \mathcal{E} = -\fft{\big(X\cdot \partial_X+Z\cdot\partial_Z\big)}{X\cdot\partial_X \big(X\cdot\partial_X+1\big)}\mathcal{D}_{\mu}^{0+}\,,\nn\\
& \mathcal{P} = \frac{2}{(X\cdot \partial_X+1) (d +X\cdot \partial_X-2) (d +2 X\cdot \partial_X-2)}\mathcal{D}_{\mu}^{+0}\,,\label{eq: ops}
\end{align}
where $-X\cdot\partial_X$ gives $\Delta$ when it acts on operators with scaling dimension $\Delta$, and $Z\cdot\partial_Z$ gives $J$ when it acts on spin-$J$ operators.

Before ending this section, we want to comment, in general, on how \eqref{eq: ops} can serve as fundamental ingredients for large- $ N $ CFT with natural power counting rules. We will show that YM and graviton amplitudes can be uplifted from flat-space to AdS by using \eqref{eq: ops}. In addition to these examples, we claim that using \eqref{eq: ops} can uplift flat-space amplitudes of effective field theories (EFT) to AdS as general large-$N$ conformal correlators, where Wilson coefficients depend on details of the conformal theory. The reason is that we find bulk derivatives can be replaced by $\mathcal{P}$ module additional terms with fewer numbers of $\mathcal{P}$. This claim implies that CFT correlators at large-$N$ limit enjoy the same power counting rules as EFTs in flat-space, which also makes manifest of the counting maps between conformal correlators and flat-space amplitudes \cite{Kravchuk:2016qvl}. Remarkably, in this way, different OPE structures can be easily distinguished. For example, three-point functions of conserved currents in generic CFT have two parity-even structures corresponding to $F^2$ and $F^3$ in AdS bulk, respectively. There was no obvious way to construct three-point structures precisely corresponding to them using embedding formalism or spin-up operators \cite{Costa:2011mg}. This is the main reason that spinning bootstrap is so hard to perform since the OPE matrix might be messy in an inappropriate basis \cite{Karateev:2018oml}. The helicity basis provides a clean way to organize the OPE matrix in CFT$_3$ \cite{Caron-Huot:2021kjy}. However, simply staring at them is still challenging to distinguish between the two structures. Now \eqref{eq: ops} makes the distinction manifest as for flat-space amplitudes! In this way, the differential representation in terms of \eqref{eq: ops} with power counting rules encoded could be useful for a clean spinning bootstrap even beyond holographic CFTs in the future. We elaborate on the discussions here in appendix \ref{app: power counting}.

\section{Construction of the differential representation}
\label{sec: diff rep}

We consider the following action for Yang-Mills theory and Einstein gravity
\be
S= \int d^{d+1}x\sqrt{g}\Big(\fft{1}{16\pi G}(R-2\Lambda)-\fft{1}{4g_{\rm YM}^2} F^a_{\mu\nu}F^{a\mu\nu}\Big)\,,\label{eq: action}
\ee
where $\Lambda=-(d-1)(d-2)/(2R_{\rm AdS}^2)$. In this paper, we usually set $R_{\rm AdS}=1$ unless we emphasize it. Our goal is to compute the four-point function for conserved currents and stress-tensors in holographic CFT that is effectively described by \eqref{eq: action}. These ``amplitudes'' can be computed by using the holographic dictionary \cite{Witten:1998qj,Gubser:1998bc}
\be
\mathcal{M}:=\langle \mathcal{O}_1\cdots \mathcal{O}_n\rangle = \big(\prod_i \fft{\delta}{\delta \mathcal{J}_i^{(0)}}\big)\left\langle e^{-S_{\rm bulk}}\right\rangle_{\rm bulk}\,,\label{eq: def AdS amp}
\ee
where $\mathcal{J}_i^{(0)}$ denotes the source as the non-normalizable mode of bulk fields, and we keep the spinning indices implicit. The variations produce the bulk-to-boundary propagators, and the remaining fields are Wick contracted by the bulk expectation value.

\subsection{Warm-up: three-point amplitudes}
\label{subsec: three-pt}

\subsubsection{Yang-Mills}

We start by looking into three-point functions as a warm-up. For YM theory, it is rather straightforward to evaluate the three-point function
\be
\mathcal{M}_{3,{\rm YM}}=g_{\rm YM}\int D^{d+1}Y f^{abc} V_{g,12}^{\mu,ab}(Y)\delta_3 A_{\mu}^c\,,\label{eq: YM three-pt vert}
\ee
where we are using the embedding AdS coordinate $Y$ \cite{Costa:2014kfa} and the shorthand notation $D^{d+1}Y:=d^{d+2}Y \delta(Y^2+1)$. For latter convenience, we explicitly write the three-point vertex function $V_{ij}^{\nu,ab}$
\begin{align}
&V_{g,12}^{\nu,ab}= \big(\nabla_\mu \delta_1 A^{\nu a} \delta_2 A^{\mu b}- \delta_1 A^{\mu a} \nabla_\mu \delta_2 A^{\nu b}\big) \nn\\
& +\fft{1}{2}\big(\delta_1 A^{\mu a}\nabla^\nu\delta_2 A_\mu^b-\nabla^\nu\delta_1 A_\mu^a\delta_2 A^{\mu b}\big) \,,\label{eq: 3-pt YM}
\end{align}
where $\delta_i A$ denotes the bulk-to-boundary propagator (in terms of the embedding space formalism \cite{Costa:2014kfa})
\be
\delta_i A^{\mu a}= \mathcal{C}_{d-1,1}\fft{2\big(X_i^\mu\, Y\cdot Z_i - Z_i^\mu \,Y\cdot X_i\big)}{\big(-2X_i\cdot Y\big)^d}\,.\label{eq: bulk-to-boundary A}
\ee
We use the standard normalization 
\be
\mathcal{C}_{\Delta,J}=\frac{\pi ^{-\frac{d}{2}} \Gamma (\Delta ) (\Delta +J-1)}{2 (\Delta -1) \Gamma \left(-\frac{d}{2}+\Delta +1\right)}\,.
\ee
Usually, it is also instructive to introduce the bulk embedding polarizations $W$ to contract the bulk indices, where $W^2=W\cdot Y\equiv 0$. Appropriate differential operators can recover the bulk indices \cite{Costa:2014kfa}. It is easy to explicitly evaluate three-point functions like \eqref{eq: 3-pt YM}. Nevertheless, in this paper, we aim to provide a differential representation that rewrites \eqref{eq: 3-pt YM} in terms of differential operators acting on scalar seeds with zero derivatives. As we claim in the last section, the bulk-to-boundary propagators shall be represented by weight-shifting operators modulo bulk coordinates. In our conventions, we find
\begin{align}
& \delta_i A_\mu=\mathcal{E}_{i,\mu}\delta_i\phi_{d-1}\,,\nn\\
& \nabla_\mu\delta_i A _\nu =\fft{d-1}{2} \Big(\mathcal{E}_{i,\nu} \mathcal{P}_{i,\mu} \delta_i\phi_{d-2} - Y_{(\mu} \mathcal{E}_{i,\nu)}\delta_i\phi_{d-1}\big)\,,\label{eq: dA rep}
\end{align}
where $\delta \phi_{\Delta}$ denotes the bulk-to-boundary propagator of scalar $\phi$ whose corresponding operator has scaling dimension $\Delta$. Aside of the second part of the second line in \eqref{eq: dA rep}, we have already observed flat-space structure by identifying $\mathcal{E}\rightarrow\epsilon$, $\mathcal{P}\rightarrow p$, and the transverse property also remains
\be
\mathcal{E}_i\cdot \mathcal{P}_i\, \delta_i \phi_{d-2} =0\,.
\ee
The overall coefficient $(d-1)/2$ seems to ruin the precise flat-space structure, but we claim this is the normalization factor that can be absorbed into the plane wave in the flat-space limit. For AdS$_4$/CFT$_3$, this normalization is precisely unit. We can then readily show \eqref{eq: 3-pt YM} can be rewritten by
\begin{align}
& \mathcal{M}_{3,{\rm YM}}= - \fft{d-1}{2}g_{\rm YM}\mathcal{T}_{\mathcal{O}}\Big((\mathcal{E}_2\cdot \mathcal{E}_3) (\mathcal{E}_1\cdot \mathcal{P}_2)W_{d-1,d-2,d-1} \nn\\
& -\big(1\leftrightarrow 2\big)+ \big(2\rightarrow1, 1\rightarrow3\big)\Big)f^{123}\,,\label{eq: YM 3-pt}
\end{align}
where $W_{\Delta_i}$ refers to the scalar contact Witten diagram with no derivatives, and $i\leftrightarrow j$ also permutes the corresponding legs for the contact Witten diagram. $\mathcal{T}_{\mathcal{O}}$ means the operator ordering, which always place $\mathcal{E}$ on the left hand side of $\mathcal{P}$ for the same point. For simplicity in this letter, we define the amplitudes by pure differential forms with contact seeds slipped off $\hat{\mathcal{M}}$. It is easy to recover the contact seeds and go from $\hat{\mathcal{M}}$ to $\mathcal{M}$ by power counting $\mathcal{P}$. By dividing appropriate normalization, \eqref{eq: YM 3-pt} is a trivial uplift from flat-space by taking $\epsilon\rightarrow \mathcal{E}$, $p\rightarrow \mathcal{P}$ followed by operator reordering. It is also straightforward to uplift another three-point vertex corresponding to the cubic term $F^3$, see appendix \ref{app: power counting}.

\subsubsection{Graviton}

For graviton three-point amplitude, we find (see also \cite{Raju:2012zs})
\be
\mathcal{M}_{3,{\rm grav}}= 4\sqrt{8\pi G}\int D^{d+1}Y V_{h,12}^{\mu\nu}(Y)\delta_3 h_{\mu\nu}\,,
\ee
where the vertex function $V_{h,12}^{\mu\nu}$ is lengthy and we leave its explicit expression to appendix \ref{app: vertex}. Similarly, $\delta_i h_{\mu\nu}$ is the bulk-to-boundary propagator for graviton, given by (we dot it into bulk embedding polarizations to keep it light)
\be
\delta_i h_{\mu\nu}W^{\mu}W^\nu = \mathcal{C}_{d,2}\fft{4\left(W\cdot X_i\, Y\cdot Z_i-W\cdot Z_i\, X_i\cdot Y\right)^2}{(-2X_i\cdot Y)^{d+2}}\,.\label{eq: bulk-to-boundary graviton}
\ee
As we promise, we find
\begin{align}
&\delta_i h_{\mu\nu} = \mathcal{E}_{i,\mu}\mathcal{E}_{i,\nu} \delta_i \phi_{d}\,,\nn\\
& \nabla_{\mu}\delta_i h_{\nu\rho} = \mathcal{E}_{i,\nu}\mathcal{E}_{i,\rho} \mathcal{P}_{i\mu}\delta_i \phi_{d-1}+\mathcal{O}(Y)\,,\nn\\
& \nabla_{\mu}\nabla_{\nu}\delta_i h_{\rho\sigma}=\mathcal{E}_{i,\rho}\mathcal{E}_{i,\sigma} \mathcal{P}_{i\mu}\mathcal{P}_{i\nu} \delta_i \phi_{d-2} + \mathcal{O}(Y,g)\,,\label{eq: ddh rep}
\end{align}
where we drop out the lengthy terms depending on bulk coordinates and metric $Y_\mu$ and $g_{\mu\nu}$. We record the complete expressions in appendix \ref{app: ddh rep and id}. As contracted, these terms $\mathcal{O}(Y,g)$ can either be annihilated or give rise to contact terms with fewer derivatives. For the three-point function, they are completely canceled, and we arrive at a precisely flat-space uplift
\be
\hat{\mathcal{M}}_{3,{\rm grav}} = \mathcal{T}_{\mathcal{O}}\Big(\mathcal{M}_{3,{\rm grav}}^{\rm flat}\Big|_{\epsilon\rightarrow\mathcal{E},p\rightarrow\mathcal{P}}\Big)\,.
\ee 

\subsubsection{The differential double copy}

Using differential operators \eqref{eq: ops}, three-point amplitudes have the same structure as in flat-space up to universal reordering. Therefore, the double copy structure at the three-point level should be straightforward. To show this, we find that although $\mathcal{P}$ itself is not conserved as the momentum, effective conservation emerges at the level of three-point amplitudes for both YM and graviton, e.g., three-point amplitudes are invariant under the following replacement
\be
\mathcal{E}_3\cdot \mathcal{P}_2 \rightarrow - \mathcal{E}_3\cdot \mathcal{P}_1\,,\quad \mathcal{P}_1\cdot \mathcal{P}_2\rightarrow 0\,.
\ee
Keeping these identities in mind, three-point amplitudes in AdS then make not much difference from flat-space, and the differential double copy is valid
\be
\hat{\mathcal{M}}_{3,\rm grav}=\fft{4}{(d-1)^2}\fft{8\sqrt{8\pi G}}{g_{\rm YM}^2}\mathcal{T}_{\mathcal{O}}\Big[(\hat{\mathcal{M}}_{3,\rm YM})^2\Big]\,.
\ee

\subsection{Four-point amplitudes}

\subsubsection{Yang-Mills}

Let us start by considering only the $s$-channel exchange diagram in YM theory
\begin{align}
\mathcal{M}_{4{\rm ex},{\rm YM}}^{(s)}=& g_{\rm YM}^2\int D^{d+1}Y_1D^{d+1}Y_2 \,f^{abc}f^{deg}\times\nn\\
 & V_{g,12}^{\mu,ab}(Y_1)\left\langle A_{\mu}^{c}(Y_1)A_{\nu}^{g}(Y_2)\right\rangle_{\rm bulk}V_{g,34}^{\nu,de}(Y_2)\,,
\end{align}
where the expectation value of the remaining bulk fields gives rise to the bulk-to-bulk propagator
\be
\left\langle A_\mu(Y_1) A_\nu(Y_2)\right\rangle:= \Pi_{g,\mu\nu}(Y_1,Y_2)\,,
\ee
which satisfies the following equation in the transverse gauge
\be
\nabla^2_{Y_1} \Pi_{g,\mu\nu}(Y_1,Y_2)= -\delta_{\mu\nu}\,\delta(Y_1-Y_2)\,.\label{eq: YM bulk-to-bulk eq}
\ee

The trick to finding the differential representation is eliminating the bulk-to-bulk propagator by the conformal Casimir operator minus its eigenvalue for the propagating field. This procedure produces effective contact diagrams that include only the bulk-to-boundary propagators. Indeed, we find
\be
\mathcal{D}_{12}^{d-1,1} V_{12}^{\mu,ab}= \nabla_{Y_1}^2 V_{12}^{\mu,ab}\,,
\ee
where 
\be
\mathcal{D}_{12}^{\Delta,J}=\mathcal{C}_{12}-\big(\Delta(\Delta-d)+J(J+d-2)\big)\,,
\ee
The conformal Casimir $\mathcal{C}_{12}$ is
\be
\mathcal{C}_{12} = - \fft{1}{2}(L_1+L_2)^2\,,\quad L_{i}^{\mu\nu}= X_{i}^{[\mu}  \partial_{X_i}^{\nu]}+Z_{i}^{[\mu}  \partial_{Z_i}^{\nu]}\,.
\ee
By integration-by-parts, we can move this bulk Laplacian to act on the bulk-to-bulk propagator and \eqref{eq: YM bulk-to-bulk eq} then reduces it to effective contact interactions (note $\mathcal{D}_{12}^{d-1,1}\equiv\mathcal{C}_{12}$)
\be
\mathcal{C}_{12}\mathcal{M}_{4{\rm ex},{\rm YM}}^{(s)} = -g_{\rm YM}^2 \int D^{d+1}Y f^{abe}f^{cde}V_{12}^{ab}(Y)\cdot V_{34}^{cd}(Y)\,,
\ee
We can then represent these effective contact terms using \eqref{eq: dA rep}. In addition to the exchanged diagram, the YM theory also provides a four-point contact diagram $A^4$. This contact diagram can be trivially uplifted from flat-space using \eqref{eq: dA rep}. Combining with permutations to include all channels, we obtain
\be
\hat{\mathcal{M}}_{4,{\rm YM}}=\fft{(d-1)^2}{4}\mathcal{T}_{\mathcal{O}}\Big(\hat{\mathcal{M}}_{4,{\rm YM}}^{\rm flat}\Big|_{\epsilon\rightarrow\mathcal{E},p\rightarrow\mathcal{P},1/s_{ij}\rightarrow 1/(2\mathcal{D}_{ij}^{d-1,1})}\Big)\,,\label{eq: uplift YM 4-pt}
\ee
where we follow \cite{Herderschee:2022ntr,Cheung:2022pdk} to define the operator $1/\mathcal{D}_{ij}$ satisfying $1/\mathcal{D}_{ij} \, \mathcal{D}_{ij}\equiv 1$, and now the operator ordering $\mathcal{T}_{\mathcal{O}}$ always keeps $1/\mathcal{D}_{ij}$ on the most left. It is worth noting that the expression of flat-space YM amplitude is not unique, as we can always use momentum conservation and generalized gauge symmetry to trade one expression for another, although they are all equivalent for on-shell amplitudes. This freedom allows for establishing the BCJ relation in the flat-space \cite{Bern:2008qj}. In equation \eqref{eq: uplift YM 4-pt}, we consistently uplift a flat-space YM amplitude obtained by following the same procedure with the same gauge and the Feynman rules in flat-space
\begin{widetext}
\begin{align}
& \hat{\mathcal{M}}_{4,{\rm YM}}^{\rm flat}=\Big[\ft{g_{\rm YM}^2}{4s_{12}} \big(2 \epsilon _1\cdot \epsilon _2\, p_1\cdot \epsilon _3\, p_3\cdot \epsilon _4-2 \epsilon _1\cdot \epsilon _2\, p_2\cdot \epsilon _3\, p_3\cdot \epsilon _4-2 \epsilon _1\cdot \epsilon _2\, p_1\cdot \epsilon _4\, p_4\cdot \epsilon _3+2 \epsilon _1\cdot \epsilon _2\, p_2\cdot \epsilon _4 p_4\cdot \epsilon _3\nn\\
&-p_1\cdot p_3\, \epsilon _3\cdot \epsilon _4\, \epsilon _1\cdot \epsilon _2 +p_1\cdot p_4 \,\epsilon _3\cdot \epsilon _4\, \epsilon _1\cdot \epsilon _2+p_2\cdot p_3\, \epsilon _3\cdot \epsilon _4\, \epsilon _1\cdot \epsilon _2-p_2\cdot p_4\, \epsilon _3\cdot \epsilon _4\, \epsilon _1\cdot \epsilon _2-4 \epsilon _1\cdot \epsilon _3\, p_1\cdot \epsilon _2\, p_3\cdot \epsilon _4\nn\\
&+4 \epsilon _1\cdot \epsilon _4\, p_1\cdot \epsilon _2\, p_4\cdot \epsilon _3 +4 \epsilon _2\cdot \epsilon _3\, p_2\cdot \epsilon _1\,  p_3\cdot \epsilon _4-4 \epsilon _2\cdot \epsilon _4\, p_2\cdot \epsilon _1\, p_4\cdot \epsilon _3+2 \epsilon _3\cdot \epsilon _4\, p_1\cdot \epsilon _2\, p_3\cdot \epsilon _1-2 \epsilon _3\cdot \epsilon _4\, p_2\cdot \epsilon _1 \,p_3\cdot \epsilon _2\nn\\
&-2 \epsilon _3\cdot \epsilon _4\, p_1\cdot \epsilon _2\, p_4\cdot \epsilon _1+2 \epsilon _3\cdot \epsilon _4\, p_2\cdot \epsilon _1\, p_4\cdot \epsilon _2\big)+ \ft{g_{\rm YM}^2}{2} \epsilon_1\cdot\epsilon_2\,\epsilon_3\cdot \epsilon_4\Big] f^{12e}f^{34e}+ \big(\text{perm}\big)\,.
\end{align}
\end{widetext}
o ensure the prescription of YM amplitudes is unambiguous and to enable the differential BCJ relation, an analogue to momentum conservation is required in AdS. However, the differential operator $\mathcal{P}$ does not provide such structures at the four-point level. Instead, we will provide an alternative differential representation using the conformal generators in the following section, which will encompass 'momentum conservation' and allow us to establish the differential BCJ relation.

\subsubsection{Graviton}

We follow the same logic for evaluating graviton amplitudes
\begin{align}
\mathcal{M}_{4{\rm ex},{\rm grav}}^{(s)}=& 16\times 8\pi G \int D^{d+1}Y_1 D^{d+1}Y_2\times \nn\\
& V_{h,12}^{\mu\nu}(Y_1) \big\langle h_{\mu\nu}(Y_1)h_{\rho\sigma}(Y_2)\big\rangle_{\rm bulk} V_{h,34}^{\rho\sigma}(Y_2)\,.\label{eq: graviton exchange diagram}
\end{align}
To correctly deal with the graviton, we must be careful about the trace part of the graviton and the vertex. We adopt the de-Donder gauge for bulk-to-bulk propagating gravitons and a meticulous analysis shows \cite{Herderschee:2022ntr, YueZhou:future} (see also appendix \ref{app: bulk-to-bulk} for more details)
\begin{align}
\mathcal{D}_{12}^{d,2}\mathcal{M}_{4{\rm ex},{\rm grav}}^{(s)}=&16\times 8\pi G \int D^{d+1}Y \times \nn\\
& V_{h,12}^{\mu\nu}(Y) P_{\mu\nu,\rho\sigma} V_{h,34}^{\rho\sigma}(Y)\,,\label{eq: D12 graviton}
\end{align}
where $P_{\mu\nu,\rho\sigma}$ is the projector precisely the same as the flat-space propagator
\be
P_{\mu\nu,\rho\sigma}=-\fft{1}{2}\Big(g_{\mu\rho}g_{\nu\sigma}+g_{\mu\sigma}g_{\rho\sigma}-\fft{2}{d-1}g_{\mu\nu}g_{\rho\sigma}\Big)\,.
\ee
The resulting contact terms can then be represented by using \eqref{eq: ddh rep} (precisely one is \eqref{eq: compete ddh rep}). To have well-defined amplitudes, we should also include four graviton contact contributions from the Einstein-Hilbert action (\eqref{eq: 4-pt ct} in appendix \ref{app: vertex}). These contributions can again be rewritten using \eqref{eq: ddh rep}. We can use many identities to eliminate all $Y$ dependences (see appendix \ref{app: ddh rep and id}). This procedure represents the graviton amplitude in terms of differential operators $\mathcal{E}$ and $\mathcal{P}$. We find that we can write down the resulting differential amplitude by uplifting the flat-space graviton amplitude plus an extra contact contribution coming from the cosmological constant term in the action $S_{\Lambda}\propto \int d^{d+1}x\sqrt{-g}\Lambda$
\be
\hat{\mathcal{M}}_{4,{\rm grav}}=& \mathcal{T}_{\mathcal{O}}\Big(\hat{\mathcal{M}}_{4,{\rm grav}}^{\rm flat}\Big|_{\epsilon\rightarrow\mathcal{E},p\rightarrow\mathcal{P},1/s_{ij}\rightarrow 1/(2\mathcal{D}_{ij}^{d,2})}\Big) \nn\\
& + \fft{1}{R_{\rm AdS}^2} \hat{\mathcal{M}}_{4,{\rm grav}}^{\rm AdS}\,.\label{eq: grav 4-pt}
\ee
Similar to YM amplitudes, $\hat{\mathcal{M}}^{\rm flat}_{4,{\rm grav}}$ is the flat-space graviton amplitude obtained using the standard Feynman techniques in flat-space. We explicitly write down $1/R_{\rm AdS}^2$ to emphasize that this term is solely contributed by the AdS term $S_{\Lambda}$ and is vanishing in the flat-space limit. This extra term is
\be
& \hat{\mathcal{M}}_{4,{\rm grav}}^{\rm AdS}=4\pi G d \Big(\big((\mathcal{E}_1\cdot\mathcal{E}_2)^2(\mathcal{E}_3\cdot\mathcal{E}_4)^2\nn\\
& -4 \mathcal{E}_1\cdot\mathcal{E}_2\,\mathcal{E}_2\cdot\mathcal{E}_3\,\mathcal{E}_3\cdot\mathcal{E}_4\,\mathcal{E}_4\cdot\mathcal{E}_1 \big)+{\rm perm} \Big)\,.\label{eq: AdS ct grav}
\ee
Under the gauge transformation $h_{\mu\nu}\rightarrow h_{\mu\nu}+\nabla_{(\mu}\xi_{\nu)}$, the action cannot be gauge invariant without $S_{\Lambda}$. For this reason, the term $\hat{\mathcal{M}}_{4,{\rm grav}}^{\rm AdS}$ in the amplitude has to exist as the consequence of the gauge invariance in the AdS bulk (which is the conservation of stress-tensors on the CFT side). This provides an idea to ``bootstrap'' stress-tensor correlators from flat-space amplitudes. To obtain the stress-tensor correlator, we can directly uplift the flat-space amplitudes and then append the enumerated crossing symmetric structures with fewer numbers of $\mathcal{P}$. The coefficients of those appended structures should be fixed by requiring the conservation of stress-tensors. However, as we will discuss in the next subsection, verifying the conservation of stress-tensors in our language is technically difficult and we did not manage to overcome this difficulty at the current stage.

\subsection{Comment on gauge invariance and conservation}

Before we end this section, we would like to discuss and comment on the relation between bulk gauge invariance and boundary conservation law using our uplift operators \eqref{eq: ops}.

From the bulk perspective, the action is invariant under the gauge transformation
\be
A_\mu\rightarrow A_\mu +\nabla_\mu\chi\,,\quad h_{\mu\nu}\rightarrow h_{\mu\nu}+\nabla_{(\mu} \xi_{\nu)}\,,
\ee
where $\chi$ is arbitrary scalar and $\xi$ vector. According to our statement that bulk derivatives can be replaced by $\mathcal{P}$ modulo bulk coordinates, we expect this gauge invariance, as in flat-space, to be represented by the invariance of boundary correlator under 
\be
\mathcal{E}_\mu\rightarrow \mathcal{E}_\mu + \# \mathcal{P}_\mu\,.\label{eq: gauge inv}
\ee
In other words, if we replace $\mathcal{E}$ by $\mathcal{P}$ (without changing the scaling dimension of the building scalar contact), the correlator should be completely vanishing
\be
\mathcal{M}\Big|_{\mathcal{E}_i\rightarrow \mathcal{P}_i}=0\,.\label{eq: conservation}
\ee
This statement reminds us of the conservation of conserved current and stress-tensor, as promised by AdS/CFT correspondence \cite{Maldacena:1997re, Witten:1998qj, Gubser:1998bc}. To show this, we should first recover the boundary tensor indices by using 
\be
\mathcal{D}_Z^\mu=\fft{d-2}{2} \partial_Z^\mu + Z\cdot\partial_Z \partial_Z^\mu -\fft{1}{2}Z^\mu \partial_Z^2\,,
\ee
then we contract one index with a boundary derivative. For the same point, we can show
\be
& \partial_X\cdot \mathcal{D}_Z\, \mathcal{E}_{\mu} = -\fft{2}{d}\mathcal{P}_\mu\,,\nn\\
& \partial_X\cdot \mathcal{D}_Z \, \mathcal{E}_{\mu} \mathcal{E}_{\nu} =-\fft{2d}{(d+2)(d-1)(d-2)}
\mathcal{P}_{\mu}\mathcal{E}_\nu\,,
\ee
where the first line is the identity for conserved currents and the second line for stress-tensors \footnote{These identities can trivially pass the strange operator $1/\mathcal{D}_{ij}$, because the Ward identity always allows us to change $(ij)$ legs to other legs that $\partial_X\cdot \mathcal{D}_Z$ does not act.}. These identities confirm that the conservation is equivalent to \eqref{eq: conservation} and is consistent with the expectation from bulk gauge invariance. We then claim we can even uplift the flat-space gauge invariance condition!

However, it is hard to impose or verify the conservation condition \eqref{eq: conservation} for differential representation. Because there are several difficulties arise after we do the replacement $\mathcal{E}\rightarrow \mathcal{P}$. Firstly, the ordering of differential operators is no longer in the operator ordering, which makes the organization messy. Secondly, it is not trivial to move differential operators to remove the poles $1/\mathcal{D}_{ij}$. In the end, we do not find conservation of $\mathcal{P}$ analogous to flat-space \footnote{Since the replacement $\mathcal{E}\rightarrow \mathcal{P}$ makes even YM amplitude not linear in $\mathcal{P}$, the simple representation using conformal generator proposed in sec \ref{sec: BCJ} does not work.}. We leave these problems to be resolved in the future works.

\section{Differential BCJ relation for YM amplitudes}
\label{sec: BCJ}

\subsection{Differential representation using conformal generators}

We have already found a differential representation for YM amplitudes in terms of the weight-shifting operators. However,  for the four-point case, we do not find the simple analogy of the flat-space momentum conservation in terms of $\mathcal{P}$. On the other hand, the conformal generators enjoy the analogy of ``momentum conservation'' because of conformal symmetry
\be
\sum_{i=1}^n L_{i}^{\mu\nu} f(X_i) \equiv 0\,,
\ee
where $f(X_i)$ is any conformal invariant function. To manifest hidden structures of YM amplitude, we propose replacing $\mathcal{P}$ with the conformal generators. Although it is not obvious, we indeed find such a replacement
\begin{align}
 \mathcal{E}_i\cdot \mathcal{P}_j \mathcal{E}_k\cdot \mathcal{P}_l \rightarrow 2\mathcal{E}_i^\mu
\mathcal{E}_k^{\nu} L_{j\mu}\,^\rho L_{l\rho\nu}\,,\quad \mathcal{P}_i\cdot \mathcal{P}_j \rightarrow -\fft{1}{2} L_i\cdot L_j\,,\label{eq: conserve replace}
\end{align}
simultaneously the scalar contact seeds are now uniform $W_{d-1,d-1,d-1,d-1}$. We now arrive at an differential representation that enjoys the ``momentum conservation'' (that we can replace $L_4$ by $-L_1-L_2-L_3$) and the transversity
\begin{align}
 \mathcal{E}_i^\mu
\mathcal{E}_k^{\nu} L_{i\mu}\,^\rho L_{l\rho\nu}= 0\,,\quad \mathcal{E}_i^\mu
\mathcal{E}_k^{\nu} L_{j\mu}\,^\rho L_{k\rho\nu}= 0\,.\label{eq: trans}
\end{align}

\subsection{Establishing the differential BCJ}

Similar to the flat-space, it is instructive to study the color-ordered amplitudes. We can easily extract the color-ordered amplitudes by recalling
\begin{align}
 f^{12e}f^{34e} &={\rm Tr}\big(T^{1}T^2 T^3 T^4\big)-{\rm Tr}\big(T^{1}T^2 T^4 T^3\big)\nn\\
& -{\rm Tr}\big(T^{1}T^3 T^4 T^2\big)+{\rm Tr}\big(T^{1}T^4 T^3 T^2\big)\,.
\end{align}
The colour-ordered amplitude $\mathcal{M}[i_1,i_2,i_3,i_4]$ is the coefficient of ${\rm Tr}\big(T^{i_1}T^{i_2}T^{i_3}T^{i_4}\big)$.

Let us take $\mathcal{M}[1234]$ as an example. We find, similar to the flat-space, we can write the color-slipped amplitude as
 \be
 \hat{\mathcal{M}}[1234]= \fft{1}{\mathcal{D}_{12}^{d-1,1}}\mathcal{N}_s - \fft{1}{\mathcal{D}_{23}^{d-1,1}}\mathcal{N}_t\,.
 \ee
As in flat-space, the differential numerator $\mathcal{N}$ is also ambiguous. For example, shifting $\mathcal{N}_s$ by $\mathcal{N}_s\rightarrow \mathcal{N}_s+{\rm const} \times \mathcal{D}_{12}^{d-1,1}$ and similarly for $\mathcal{N}_t$ doesn't change the amplitudes. It is not hard to find such numerators satisfying the following permutation properties
\be
\mathcal{N}_s\big|_{2\leftrightarrow4}=-\mathcal{N}_t\,,\quad \mathcal{N}_{s}\big|_{2\rightarrow 4,3\rightarrow2,4\rightarrow3}= \mathcal{N}_t\,,
\ee
which could be obtained by uplifting flat-space numerators
\be
\mathcal{N}_s = \mathcal{T}_{\mathcal{O}}\Big(n_s^{\rm flat}\Big|_{\epsilon\rightarrow\mathcal{E},p\rightarrow\mathcal{P}}\Big)\,,
\ee
followed by replacement \eqref{eq: conserve replace}. Nevertheless, even though we uplift the flat BCJ numerators that satisfy the kinematic Jacobi relation \cite{Bern:2008qj}
\be
n_s+n_t+n_u=0\,,
\ee
it still does not guarantee that the differential numerators satisfy these kinematic Jacobi relations. The culprit is the ordering of differential operators; more specifically, terms like $f(\mathcal{E})L_i\cdot L_j$ do not manifestly cancel the denominator $\mathcal{D}^{d-1,1}$. To resolve this problem, we should move $L_i\cdot L_j$ to the most left. We can easily do this by noting
\be
&\mathcal{E}_{i_1}\cdot \mathcal{E}_{i_2}\,\mathcal{E}_{i_3}\cdot \mathcal{E}_{i_4}\, L_{i_1}\cdot L_{i_3}=\mathcal{D}_{i_1i_3}^{d-1,1} \mathcal{E}_{i_1}\cdot \mathcal{E}_{i_2}\,\mathcal{E}_{i_3}\cdot \mathcal{E}_{i_4}\nn\\
& - 2\big(\mathcal{E}_{i_2}\cdot \mathcal{E}_{i_4}\,\mathcal{E}_{i_1}\cdot \mathcal{E}_{i_3}-\mathcal{E}_{i_2}\cdot \mathcal{E}_{i_3}\,\mathcal{E}_{i_1}\cdot \mathcal{E}_{i_4}\big)\,.
\ee
It turns out that additional terms such as the second line above would generally cancel out in the final expression, and we then trivially move $L_i\cdot L_j$ to the most left as $\mathcal{D}_{ij}^{d-1,1}$. After this operation, we can then use the Ward identity to rewrite $L_4=-L_1-L_2-L_3$ followed by the transversity \eqref{eq: trans} to eliminate unwanted terms. We also have to show
\be
\big(\mathcal{D}^{d-1,1}_{12}+\mathcal{D}^{d-1,1}_{13}+\mathcal{D}^{d-1,1}_{23}\big)f_{1,d-1}(X_i)=0\,,
\ee
where $f_{1,d-1}$ is any conformal invariant function with spin weights $J=1$ and scaling weights $\Delta=d-1$. This statement is equivalent to $L_i^2 f_{1,d-1}=0$, which can be easily proved by acting $L_i^2$ on shadow representation \cite{SimmonsDuffin:2012uy,Karateev:2018oml} of any such function $f_{1,d-1}$
\be
f_{1,d-1}=& \sum_J \int d\Delta I(\Delta,J)\times\nn\\
& \int D^dX_5 \langle V_1 V_2 \mathcal{O}_{\Delta,J}(X_5)\rangle \langle \tilde{\mathcal{O}}_{d-\Delta,J}(X_5)V_3 V_4\rangle\,,
\ee
where $V_i$ is used to denote the conserved current operator in CFT, while $\mathcal{O}_{\Delta,J}$ is an arbitrary spin-$J$ operator with scaling dimension $\Delta$ that can appear. Taking all of these into account, we can then prove the differential kinematic Jacobi identity
\be
\mathcal{N}_s+\mathcal{N}_t+\mathcal{N}_u=0\,.\label{eq: diff BCJ num}
\ee

Following the same operations described above, we can readily prove the differential BCJ relation
\be
\mathcal{D}_{12}^{d-1,1}\hat{\mathcal{M}}[1234]-\mathcal{D}_{13}^{d-1,1}\hat{\mathcal{M}}[1324]
=0\,.
\ee

\subsection{Comment on four-point double copy}

The uplift of the BCJ numerators \eqref{eq: diff BCJ num} strongly suggests that there should be a differential double copy relation up to terms with less number of momenta suppressed by the AdS radius, namely,
\be
\hat{\mathcal{M}}_{4,{\rm grav}} \propto \mathcal{T}_{\mathcal{O}}\Big(\fft{1}{\mathcal{D}_{12}^{d,2}}\mathcal{N}_s^2+\fft{1}{\mathcal{D}_{23}^{d,2}}\mathcal{N}_t^2+\fft{1}{\mathcal{D}_{24}^{d,2}}\mathcal{N}_u^2\Big) +\hat{\mathcal{M}}_{\rm res}\,.\label{eq: double copy contact}
\ee
However, it is hard to find the remaining term $\hat{\mathcal{M}}_{\rm res}$ and prove this proposal. The most important reason is that the momentum conservation is built into the double copy relation, but we do not manage to find a clean way to replace $\mathcal{P}$ in \eqref{eq: grav 4-pt} by $L$. There are large redundancies to rewrite \eqref{eq: grav 4-pt} in terms of conformal generators. It is thus difficult to locate a nice minimum basis that allows us to prove \eqref{eq: double copy contact} by figuring out what is $\hat{\mathcal{M}}_{\rm res}$. Another way to explore \eqref{eq: double copy contact} might be generalizing the algorithm in \cite{Li:2023azu} that translates the differential representation to final amplitudes in the Mellin space. The resulting Mellin amplitudes \cite{Mack:2009mi, Penedones:2010ue} may help explicitly verify the relation and fill in the missing corner $\hat{\mathcal{M}}_{\rm res}$. We can also hope to completely determine $\hat{\mathcal{M}}_{\rm res}$ by enumerating all possible structures with less number of $\mathcal{P}$ and requiring the conservation of stress-tensors in the stress-tensor correlator $\hat{\mathcal{M}}_{4,{\rm grav}}$. We leave this interesting question for future studies.

\section{Summary}
\label{sec: summary}

We proposed the differential representation for tree-level gluon and graviton scattering from YM and Einstein gravity in AdS. The essential differential operators are proportional to dimension-up and spin-up weight-shifting operators. They provide a natural scheme for organizing (A)dS amplitudes and large-$N$ conformal correlators by counting the number of $\mathcal{P}$, where the hierarchy of different structures is made manifest, as we explain in appendix \ref{app: power counting}. Using these differential operators, three-point and four-point amplitudes in AdS are straightforwardly uplifted from flat-space cousins. For three-point amplitudes, such an uplift makes the double copy relation straightforward. At the four-point level, we find a different differential representation for YM amplitudes by using spin-up weight-shifting operators and the conformal generators, for which differential BCJ numerators can be uplifted from flat-space ones, building differential BCJ relations. The differential BCJ numerators follow the kinematic Jacobi identity. We could then argue that the double copy structure for the four-point function should be valid up to the remaining terms with a fewer number of weight-shifting operators $\mathcal{P}$. It would be interesting to make connections between our findings and the similar structures in momentum space \cite{Armstrong:2020woi,Albayrak:2020fyp,Lee:2022fgr,Farrow:2018yni,Farrow:2018yni,Armstrong:2022mfr,Armstrong:2022csc} or Mellin space \cite{Alday:2021odx,Zhou:2021gnu,Bissi:2022wuh} (by generalizing to supersymmetric theories \cite{Roehrig:2020kck}). These connections, as analytically continued to the dS space \cite{Sleight:2020obc,Sleight:2021plv}, could improve the understanding of cosmological correlators by following the lines of, e.g., \cite{Baumann:2022jpr,Sleight:2019mgd,Arkani-Hamed:2018kmz,Baumann:2019oyu,Baumann:2020ksv,Baumann:2020dch}.

This paper is the first step toward revealing the hidden structures of spinning correlators. Most importantly, our surprising findings rely on more or less guessing work. It is thus crucial to develop a more systematic way for uplifting by using \eqref{eq: ops}, and relevant operators, similar to the scalar case \cite{Eberhardt:2020ewh,Cheung:2022pdk}. Besides, the ordering of differential operators and non-conservation of operators $\mathcal{P}$ prevent one from proving or imposing the conservation for current and stress-tensor operators. One possible way to resolve this problem is to carefully think about algebra that \eqref{eq: ops} may form together with other differential operators (such as emergent ${\rm SO}(5,5)$ algebra for bispinor representation of AdS$_4$/CFT$_3$ \cite{Binder:2020raz}). Besides, the uplift from flat-space convinces us there should also exist the Parke-Taylor formula, and we believe the bispinor formalism \cite{Binder:2020raz} would be the correct tool. As these challenges are overcome, we believe the four-point double copy in AdS can be precisely established. 

It would also be interesting to understand why this differential representation manifests the flat-space limit by detailed investigation of the Inönü-Wigner contraction of the conformal group, and its representations \cite{Goncalves:2014rfa}. This exploration can help understand many aspects of S-matrix as the flat-space limit of conformal correlators, following the lines of \cite{Komatsu:2020sag,vanRees:2022itk,Caron-Huot:2021enk,Duary:2022pyv,Duary:2022afn}.

Ultimately, we want to emphasize that the differential representation might help bootstrap holographic CFTs beyond the scope of the Lorentzian inversion formula \cite{Caron-Huot:2017vep,Simmons-Duffin:2017nub,Kravchuk:2018htv}. The Lorentzian inversion formula does not work well for spin-zero trajectories, while the differential representation precisely captures the contact terms with complete OPE data built into the numerators.

\begin{acknowledgments}
We thank Simon Caron-Huot, Julio Parra-Martinez, Jiajie Mei, Joao Penedones and David Simmons-Duffin for discussions. We thank Yi Wang for the hospitality during my visit to HKUST Jockey Club Institute for Advanced Study. YZL is supported by the Simons Foundation through the Simons Collaboration on the Nonperturbative Bootstrap.
\end{acknowledgments}

\bibliographystyle{apsrev4-1}
\bibliography{ref.bib}

\begin{thebibliography}{74}%
\makeatletter
\providecommand \@ifxundefined [1]{%
 \@ifx{#1\undefined}
}%
\providecommand \@ifnum [1]{%
 \ifnum #1\expandafter \@firstoftwo
 \else \expandafter \@secondoftwo
 \fi
}%
\providecommand \@ifx [1]{%
 \ifx #1\expandafter \@firstoftwo
 \else \expandafter \@secondoftwo
 \fi
}%
\providecommand \natexlab [1]{#1}%
\providecommand \enquote  [1]{``#1''}%
\providecommand \bibnamefont  [1]{#1}%
\providecommand \bibfnamefont [1]{#1}%
\providecommand \citenamefont [1]{#1}%
\providecommand \href@noop [0]{\@secondoftwo}%
\providecommand \href [0]{\begingroup \@sanitize@url \@href}%
\providecommand \@href[1]{\@@startlink{#1}\@@href}%
\providecommand \@@href[1]{\endgroup#1\@@endlink}%
\providecommand \@sanitize@url [0]{\catcode `\\12\catcode `\$12\catcode
  `\&12\catcode `\#12\catcode `\^12\catcode `\_12\catcode `\%12\relax}%
\providecommand \@@startlink[1]{}%
\providecommand \@@endlink[0]{}%
\providecommand \url  [0]{\begingroup\@sanitize@url \@url }%
\providecommand \@url [1]{\endgroup\@href {#1}{\urlprefix }}%
\providecommand \urlprefix  [0]{URL }%
\providecommand \Eprint [0]{\href }%
\providecommand \doibase [0]{http://dx.doi.org/}%
\providecommand \selectlanguage [0]{\@gobble}%
\providecommand \bibinfo  [0]{\@secondoftwo}%
\providecommand \bibfield  [0]{\@secondoftwo}%
\providecommand \translation [1]{[#1]}%
\providecommand \BibitemOpen [0]{}%
\providecommand \bibitemStop [0]{}%
\providecommand \bibitemNoStop [0]{.\EOS\space}%
\providecommand \EOS [0]{\spacefactor3000\relax}%
\providecommand \BibitemShut  [1]{\csname bibitem#1\endcsname}%
\let\auto@bib@innerbib\@empty
\bibitem [{\citenamefont {Weinberg}(2005)}]{Weinberg:1995mt}%
  \BibitemOpen
  \bibfield  {author} {\bibinfo {author} {\bibfnamefont {S.}~\bibnamefont
  {Weinberg}},\ }\href@noop {} {\emph {\bibinfo {title} {{The Quantum theory of
  fields. Vol. 1: Foundations}}}}\ (\bibinfo  {publisher} {Cambridge University
  Press},\ \bibinfo {year} {2005})\BibitemShut {NoStop}%
\bibitem [{\citenamefont {Bern}\ \emph {et~al.}(2008)\citenamefont {Bern},
  \citenamefont {Carrasco},\ and\ \citenamefont {Johansson}}]{Bern:2008qj}%
  \BibitemOpen
  \bibfield  {author} {\bibinfo {author} {\bibfnamefont {Z.}~\bibnamefont
  {Bern}}, \bibinfo {author} {\bibfnamefont {J.~J.~M.}\ \bibnamefont
  {Carrasco}}, \ and\ \bibinfo {author} {\bibfnamefont {H.}~\bibnamefont
  {Johansson}},\ }\href {\doibase 10.1103/PhysRevD.78.085011} {\bibfield
  {journal} {\bibinfo  {journal} {Phys. Rev. D}\ }\textbf {\bibinfo {volume}
  {78}},\ \bibinfo {pages} {085011} (\bibinfo {year} {2008})},\ \Eprint
  {http://arxiv.org/abs/0805.3993} {arXiv:0805.3993 [hep-ph]} \BibitemShut
  {NoStop}%
\bibitem [{\citenamefont {Kawai}\ \emph {et~al.}(1986)\citenamefont {Kawai},
  \citenamefont {Lewellen},\ and\ \citenamefont {Tye}}]{KAWAI19861}%
  \BibitemOpen
  \bibfield  {author} {\bibinfo {author} {\bibfnamefont {H.}~\bibnamefont
  {Kawai}}, \bibinfo {author} {\bibfnamefont {D.}~\bibnamefont {Lewellen}}, \
  and\ \bibinfo {author} {\bibfnamefont {S.-H.}\ \bibnamefont {Tye}},\ }\href
  {\doibase https://doi.org/10.1016/0550-3213(86)90362-7} {\bibfield  {journal}
  {\bibinfo  {journal} {Nuclear Physics B}\ }\textbf {\bibinfo {volume}
  {269}},\ \bibinfo {pages} {1} (\bibinfo {year} {1986})}\BibitemShut {NoStop}%
\bibitem [{\citenamefont {Bern}\ \emph {et~al.}(2010)\citenamefont {Bern},
  \citenamefont {Carrasco},\ and\ \citenamefont {Johansson}}]{Bern:2010ue}%
  \BibitemOpen
  \bibfield  {author} {\bibinfo {author} {\bibfnamefont {Z.}~\bibnamefont
  {Bern}}, \bibinfo {author} {\bibfnamefont {J.~J.~M.}\ \bibnamefont
  {Carrasco}}, \ and\ \bibinfo {author} {\bibfnamefont {H.}~\bibnamefont
  {Johansson}},\ }\href {\doibase 10.1103/PhysRevLett.105.061602} {\bibfield
  {journal} {\bibinfo  {journal} {Phys. Rev. Lett.}\ }\textbf {\bibinfo
  {volume} {105}},\ \bibinfo {pages} {061602} (\bibinfo {year} {2010})},\
  \Eprint {http://arxiv.org/abs/1004.0476} {arXiv:1004.0476 [hep-th]}
  \BibitemShut {NoStop}%
\bibitem [{\citenamefont {Bern}\ \emph {et~al.}(2019)\citenamefont {Bern},
  \citenamefont {Carrasco}, \citenamefont {Chiodaroli}, \citenamefont
  {Johansson},\ and\ \citenamefont {Roiban}}]{Bern:2019prr}%
  \BibitemOpen
  \bibfield  {author} {\bibinfo {author} {\bibfnamefont {Z.}~\bibnamefont
  {Bern}}, \bibinfo {author} {\bibfnamefont {J.~J.}\ \bibnamefont {Carrasco}},
  \bibinfo {author} {\bibfnamefont {M.}~\bibnamefont {Chiodaroli}}, \bibinfo
  {author} {\bibfnamefont {H.}~\bibnamefont {Johansson}}, \ and\ \bibinfo
  {author} {\bibfnamefont {R.}~\bibnamefont {Roiban}},\ }\href@noop {} {\
  (\bibinfo {year} {2019})},\ \Eprint {http://arxiv.org/abs/1909.01358}
  {arXiv:1909.01358 [hep-th]} \BibitemShut {NoStop}%
\bibitem [{\citenamefont {Maldacena}(1998)}]{Maldacena:1997re}%
  \BibitemOpen
  \bibfield  {author} {\bibinfo {author} {\bibfnamefont {J.~M.}\ \bibnamefont
  {Maldacena}},\ }\href {\doibase 10.1023/A:1026654312961} {\bibfield
  {journal} {\bibinfo  {journal} {Adv. Theor. Math. Phys.}\ }\textbf {\bibinfo
  {volume} {2}},\ \bibinfo {pages} {231} (\bibinfo {year} {1998})},\ \Eprint
  {http://arxiv.org/abs/hep-th/9711200} {arXiv:hep-th/9711200} \BibitemShut
  {NoStop}%
\bibitem [{Note1()}]{Note1}%
  \BibitemOpen
  \bibinfo {note} {For dS, although it turns out its essential structures are
  similar to AdS and analytic continuation exists to go from one to another
  \cite {Sleight:2020obc}, the unitarity of dS scattering is not a standard
  concept from CFT perspective \cite {Hogervorst:2021uvp}.}\BibitemShut {Stop}%
\bibitem [{\citenamefont {Hartman}\ \emph {et~al.}(2022)\citenamefont
  {Hartman}, \citenamefont {Mazac}, \citenamefont {Simmons-Duffin},\ and\
  \citenamefont {Zhiboedov}}]{Hartman:2022zik}%
  \BibitemOpen
  \bibfield  {author} {\bibinfo {author} {\bibfnamefont {T.}~\bibnamefont
  {Hartman}}, \bibinfo {author} {\bibfnamefont {D.}~\bibnamefont {Mazac}},
  \bibinfo {author} {\bibfnamefont {D.}~\bibnamefont {Simmons-Duffin}}, \ and\
  \bibinfo {author} {\bibfnamefont {A.}~\bibnamefont {Zhiboedov}},\ }in\
  \href@noop {} {\emph {\bibinfo {booktitle} {{2022 Snowmass Summer Study}}}}\
  (\bibinfo {year} {2022})\ \Eprint {http://arxiv.org/abs/2202.11012}
  {arXiv:2202.11012 [hep-th]} \BibitemShut {NoStop}%
\bibitem [{\citenamefont {Bissi}\ \emph
  {et~al.}(2022{\natexlab{a}})\citenamefont {Bissi}, \citenamefont {Sinha},\
  and\ \citenamefont {Zhou}}]{Bissi:2022mrs}%
  \BibitemOpen
  \bibfield  {author} {\bibinfo {author} {\bibfnamefont {A.}~\bibnamefont
  {Bissi}}, \bibinfo {author} {\bibfnamefont {A.}~\bibnamefont {Sinha}}, \ and\
  \bibinfo {author} {\bibfnamefont {X.}~\bibnamefont {Zhou}},\ }\href {\doibase
  10.1016/j.physrep.2022.09.004} {\bibfield  {journal} {\bibinfo  {journal}
  {Phys. Rept.}\ }\textbf {\bibinfo {volume} {991}},\ \bibinfo {pages} {1}
  (\bibinfo {year} {2022}{\natexlab{a}})},\ \Eprint
  {http://arxiv.org/abs/2202.08475} {arXiv:2202.08475 [hep-th]} \BibitemShut
  {NoStop}%
\bibitem [{\citenamefont {Okuda}\ and\ \citenamefont
  {Penedones}(2011)}]{Okuda:2010ym}%
  \BibitemOpen
  \bibfield  {author} {\bibinfo {author} {\bibfnamefont {T.}~\bibnamefont
  {Okuda}}\ and\ \bibinfo {author} {\bibfnamefont {J.}~\bibnamefont
  {Penedones}},\ }\href {\doibase 10.1103/PhysRevD.83.086001} {\bibfield
  {journal} {\bibinfo  {journal} {Phys. Rev. D}\ }\textbf {\bibinfo {volume}
  {83}},\ \bibinfo {pages} {086001} (\bibinfo {year} {2011})},\ \Eprint
  {http://arxiv.org/abs/1002.2641} {arXiv:1002.2641 [hep-th]} \BibitemShut
  {NoStop}%
\bibitem [{\citenamefont {Penedones}(2011)}]{Penedones:2010ue}%
  \BibitemOpen
  \bibfield  {author} {\bibinfo {author} {\bibfnamefont {J.}~\bibnamefont
  {Penedones}},\ }\href {\doibase 10.1007/JHEP03(2011)025} {\bibfield
  {journal} {\bibinfo  {journal} {JHEP}\ }\textbf {\bibinfo {volume} {03}},\
  \bibinfo {pages} {025} (\bibinfo {year} {2011})},\ \Eprint
  {http://arxiv.org/abs/1011.1485} {arXiv:1011.1485 [hep-th]} \BibitemShut
  {NoStop}%
\bibitem [{\citenamefont {Fitzpatrick}\ and\ \citenamefont
  {Kaplan}(2012)}]{Fitzpatrick:2011hu}%
  \BibitemOpen
  \bibfield  {author} {\bibinfo {author} {\bibfnamefont {A.~L.}\ \bibnamefont
  {Fitzpatrick}}\ and\ \bibinfo {author} {\bibfnamefont {J.}~\bibnamefont
  {Kaplan}},\ }\href {\doibase 10.1007/JHEP10(2012)127} {\bibfield  {journal}
  {\bibinfo  {journal} {JHEP}\ }\textbf {\bibinfo {volume} {10}},\ \bibinfo
  {pages} {127} (\bibinfo {year} {2012})},\ \Eprint
  {http://arxiv.org/abs/1111.6972} {arXiv:1111.6972 [hep-th]} \BibitemShut
  {NoStop}%
\bibitem [{\citenamefont {Maldacena}\ \emph {et~al.}(2017)\citenamefont
  {Maldacena}, \citenamefont {Simmons-Duffin},\ and\ \citenamefont
  {Zhiboedov}}]{Maldacena:2015iua}%
  \BibitemOpen
  \bibfield  {author} {\bibinfo {author} {\bibfnamefont {J.}~\bibnamefont
  {Maldacena}}, \bibinfo {author} {\bibfnamefont {D.}~\bibnamefont
  {Simmons-Duffin}}, \ and\ \bibinfo {author} {\bibfnamefont {A.}~\bibnamefont
  {Zhiboedov}},\ }\href {\doibase 10.1007/JHEP01(2017)013} {\bibfield
  {journal} {\bibinfo  {journal} {JHEP}\ }\textbf {\bibinfo {volume} {01}},\
  \bibinfo {pages} {013} (\bibinfo {year} {2017})},\ \Eprint
  {http://arxiv.org/abs/1509.03612} {arXiv:1509.03612 [hep-th]} \BibitemShut
  {NoStop}%
\bibitem [{\citenamefont {Raju}(2012{\natexlab{a}})}]{Raju:2012zr}%
  \BibitemOpen
  \bibfield  {author} {\bibinfo {author} {\bibfnamefont {S.}~\bibnamefont
  {Raju}},\ }\href {\doibase 10.1103/PhysRevD.85.126009} {\bibfield  {journal}
  {\bibinfo  {journal} {Phys. Rev. D}\ }\textbf {\bibinfo {volume} {85}},\
  \bibinfo {pages} {126009} (\bibinfo {year} {2012}{\natexlab{a}})},\ \Eprint
  {http://arxiv.org/abs/1201.6449} {arXiv:1201.6449 [hep-th]} \BibitemShut
  {NoStop}%
\bibitem [{\citenamefont {Paulos}\ \emph {et~al.}(2017)\citenamefont {Paulos},
  \citenamefont {Penedones}, \citenamefont {Toledo}, \citenamefont {van Rees},\
  and\ \citenamefont {Vieira}}]{Paulos:2016fap}%
  \BibitemOpen
  \bibfield  {author} {\bibinfo {author} {\bibfnamefont {M.~F.}\ \bibnamefont
  {Paulos}}, \bibinfo {author} {\bibfnamefont {J.}~\bibnamefont {Penedones}},
  \bibinfo {author} {\bibfnamefont {J.}~\bibnamefont {Toledo}}, \bibinfo
  {author} {\bibfnamefont {B.~C.}\ \bibnamefont {van Rees}}, \ and\ \bibinfo
  {author} {\bibfnamefont {P.}~\bibnamefont {Vieira}},\ }\href {\doibase
  10.1007/JHEP11(2017)133} {\bibfield  {journal} {\bibinfo  {journal} {JHEP}\
  }\textbf {\bibinfo {volume} {11}},\ \bibinfo {pages} {133} (\bibinfo {year}
  {2017})},\ \Eprint {http://arxiv.org/abs/1607.06109} {arXiv:1607.06109
  [hep-th]} \BibitemShut {NoStop}%
\bibitem [{\citenamefont {Komatsu}\ \emph {et~al.}(2020)\citenamefont
  {Komatsu}, \citenamefont {Paulos}, \citenamefont {Van~Rees},\ and\
  \citenamefont {Zhao}}]{Komatsu:2020sag}%
  \BibitemOpen
  \bibfield  {author} {\bibinfo {author} {\bibfnamefont {S.}~\bibnamefont
  {Komatsu}}, \bibinfo {author} {\bibfnamefont {M.~F.}\ \bibnamefont {Paulos}},
  \bibinfo {author} {\bibfnamefont {B.~C.}\ \bibnamefont {Van~Rees}}, \ and\
  \bibinfo {author} {\bibfnamefont {X.}~\bibnamefont {Zhao}},\ }\href {\doibase
  10.1007/JHEP11(2020)046} {\bibfield  {journal} {\bibinfo  {journal} {JHEP}\
  }\textbf {\bibinfo {volume} {11}},\ \bibinfo {pages} {046} (\bibinfo {year}
  {2020})},\ \Eprint {http://arxiv.org/abs/2007.13745} {arXiv:2007.13745
  [hep-th]} \BibitemShut {NoStop}%
\bibitem [{\citenamefont {Hijano}(2019)}]{Hijano:2019qmi}%
  \BibitemOpen
  \bibfield  {author} {\bibinfo {author} {\bibfnamefont {E.}~\bibnamefont
  {Hijano}},\ }\href {\doibase 10.1007/JHEP07(2019)132} {\bibfield  {journal}
  {\bibinfo  {journal} {JHEP}\ }\textbf {\bibinfo {volume} {07}},\ \bibinfo
  {pages} {132} (\bibinfo {year} {2019})},\ \Eprint
  {http://arxiv.org/abs/1905.02729} {arXiv:1905.02729 [hep-th]} \BibitemShut
  {NoStop}%
\bibitem [{\citenamefont {Li}(2021)}]{Li:2021snj}%
  \BibitemOpen
  \bibfield  {author} {\bibinfo {author} {\bibfnamefont {Y.-Z.}\ \bibnamefont
  {Li}},\ }\href {\doibase 10.1007/JHEP09(2021)027} {\bibfield  {journal}
  {\bibinfo  {journal} {JHEP}\ }\textbf {\bibinfo {volume} {09}},\ \bibinfo
  {pages} {027} (\bibinfo {year} {2021})},\ \Eprint
  {http://arxiv.org/abs/2106.04606} {arXiv:2106.04606 [hep-th]} \BibitemShut
  {NoStop}%
\bibitem [{Note2()}]{Note2}%
  \BibitemOpen
  \bibinfo {note} {However, there exist exemptions for special analytic regimes
  that are not well-understood yet, see \cite {Komatsu:2020sag} for recent
  explorations.}\BibitemShut {Stop}%
\bibitem [{\citenamefont {Bagchi}\ \emph {et~al.}(2016)\citenamefont {Bagchi},
  \citenamefont {Basu}, \citenamefont {Kakkar},\ and\ \citenamefont
  {Mehra}}]{Bagchi:2016bcd}%
  \BibitemOpen
  \bibfield  {author} {\bibinfo {author} {\bibfnamefont {A.}~\bibnamefont
  {Bagchi}}, \bibinfo {author} {\bibfnamefont {R.}~\bibnamefont {Basu}},
  \bibinfo {author} {\bibfnamefont {A.}~\bibnamefont {Kakkar}}, \ and\ \bibinfo
  {author} {\bibfnamefont {A.}~\bibnamefont {Mehra}},\ }\href {\doibase
  10.1007/JHEP12(2016)147} {\bibfield  {journal} {\bibinfo  {journal} {JHEP}\
  }\textbf {\bibinfo {volume} {12}},\ \bibinfo {pages} {147} (\bibinfo {year}
  {2016})},\ \Eprint {http://arxiv.org/abs/1609.06203} {arXiv:1609.06203
  [hep-th]} \BibitemShut {NoStop}%
\bibitem [{\citenamefont {Eberhardt}\ \emph {et~al.}(2020)\citenamefont
  {Eberhardt}, \citenamefont {Komatsu},\ and\ \citenamefont
  {Mizera}}]{Eberhardt:2020ewh}%
  \BibitemOpen
  \bibfield  {author} {\bibinfo {author} {\bibfnamefont {L.}~\bibnamefont
  {Eberhardt}}, \bibinfo {author} {\bibfnamefont {S.}~\bibnamefont {Komatsu}},
  \ and\ \bibinfo {author} {\bibfnamefont {S.}~\bibnamefont {Mizera}},\ }\href
  {\doibase 10.1007/JHEP11(2020)158} {\bibfield  {journal} {\bibinfo  {journal}
  {JHEP}\ }\textbf {\bibinfo {volume} {11}},\ \bibinfo {pages} {158} (\bibinfo
  {year} {2020})},\ \Eprint {http://arxiv.org/abs/2007.06574} {arXiv:2007.06574
  [hep-th]} \BibitemShut {NoStop}%
\bibitem [{\citenamefont {Roehrig}\ and\ \citenamefont
  {Skinner}(2022)}]{Roehrig:2020kck}%
  \BibitemOpen
  \bibfield  {author} {\bibinfo {author} {\bibfnamefont {K.}~\bibnamefont
  {Roehrig}}\ and\ \bibinfo {author} {\bibfnamefont {D.}~\bibnamefont
  {Skinner}},\ }\href {\doibase 10.1007/JHEP02(2022)073} {\bibfield  {journal}
  {\bibinfo  {journal} {JHEP}\ }\textbf {\bibinfo {volume} {02}},\ \bibinfo
  {pages} {073} (\bibinfo {year} {2022})},\ \Eprint
  {http://arxiv.org/abs/2007.07234} {arXiv:2007.07234 [hep-th]} \BibitemShut
  {NoStop}%
\bibitem [{\citenamefont {Diwakar}\ \emph {et~al.}(2021)\citenamefont
  {Diwakar}, \citenamefont {Herderschee}, \citenamefont {Roiban},\ and\
  \citenamefont {Teng}}]{Diwakar:2021juk}%
  \BibitemOpen
  \bibfield  {author} {\bibinfo {author} {\bibfnamefont {P.}~\bibnamefont
  {Diwakar}}, \bibinfo {author} {\bibfnamefont {A.}~\bibnamefont
  {Herderschee}}, \bibinfo {author} {\bibfnamefont {R.}~\bibnamefont {Roiban}},
  \ and\ \bibinfo {author} {\bibfnamefont {F.}~\bibnamefont {Teng}},\ }\href
  {\doibase 10.1007/JHEP10(2021)141} {\bibfield  {journal} {\bibinfo  {journal}
  {JHEP}\ }\textbf {\bibinfo {volume} {10}},\ \bibinfo {pages} {141} (\bibinfo
  {year} {2021})},\ \Eprint {http://arxiv.org/abs/2106.10822} {arXiv:2106.10822
  [hep-th]} \BibitemShut {NoStop}%
\bibitem [{\citenamefont {Herderschee}\ \emph {et~al.}(2022)\citenamefont
  {Herderschee}, \citenamefont {Roiban},\ and\ \citenamefont
  {Teng}}]{Herderschee:2022ntr}%
  \BibitemOpen
  \bibfield  {author} {\bibinfo {author} {\bibfnamefont {A.}~\bibnamefont
  {Herderschee}}, \bibinfo {author} {\bibfnamefont {R.}~\bibnamefont {Roiban}},
  \ and\ \bibinfo {author} {\bibfnamefont {F.}~\bibnamefont {Teng}},\ }\href
  {\doibase 10.1007/JHEP05(2022)026} {\bibfield  {journal} {\bibinfo  {journal}
  {JHEP}\ }\textbf {\bibinfo {volume} {05}},\ \bibinfo {pages} {026} (\bibinfo
  {year} {2022})},\ \Eprint {http://arxiv.org/abs/2201.05067} {arXiv:2201.05067
  [hep-th]} \BibitemShut {NoStop}%
\bibitem [{\citenamefont {Cheung}\ \emph {et~al.}(2022)\citenamefont {Cheung},
  \citenamefont {Parra-Martinez},\ and\ \citenamefont
  {Sivaramakrishnan}}]{Cheung:2022pdk}%
  \BibitemOpen
  \bibfield  {author} {\bibinfo {author} {\bibfnamefont {C.}~\bibnamefont
  {Cheung}}, \bibinfo {author} {\bibfnamefont {J.}~\bibnamefont
  {Parra-Martinez}}, \ and\ \bibinfo {author} {\bibfnamefont {A.}~\bibnamefont
  {Sivaramakrishnan}},\ }\href {\doibase 10.1007/JHEP05(2022)027} {\bibfield
  {journal} {\bibinfo  {journal} {JHEP}\ }\textbf {\bibinfo {volume} {05}},\
  \bibinfo {pages} {027} (\bibinfo {year} {2022})},\ \Eprint
  {http://arxiv.org/abs/2201.05147} {arXiv:2201.05147 [hep-th]} \BibitemShut
  {NoStop}%
\bibitem [{\citenamefont {Gomez}\ \emph {et~al.}(2021)\citenamefont {Gomez},
  \citenamefont {Jusinskas},\ and\ \citenamefont {Lipstein}}]{Gomez:2021qfd}%
  \BibitemOpen
  \bibfield  {author} {\bibinfo {author} {\bibfnamefont {H.}~\bibnamefont
  {Gomez}}, \bibinfo {author} {\bibfnamefont {R.~L.}\ \bibnamefont
  {Jusinskas}}, \ and\ \bibinfo {author} {\bibfnamefont {A.}~\bibnamefont
  {Lipstein}},\ }\href {\doibase 10.1103/PhysRevLett.127.251604} {\bibfield
  {journal} {\bibinfo  {journal} {Phys. Rev. Lett.}\ }\textbf {\bibinfo
  {volume} {127}},\ \bibinfo {pages} {251604} (\bibinfo {year} {2021})},\
  \Eprint {http://arxiv.org/abs/2106.11903} {arXiv:2106.11903 [hep-th]}
  \BibitemShut {NoStop}%
\bibitem [{\citenamefont {Gomez}\ \emph {et~al.}(2022)\citenamefont {Gomez},
  \citenamefont {Lipinski~Jusinskas},\ and\ \citenamefont
  {Lipstein}}]{Gomez:2021ujt}%
  \BibitemOpen
  \bibfield  {author} {\bibinfo {author} {\bibfnamefont {H.}~\bibnamefont
  {Gomez}}, \bibinfo {author} {\bibfnamefont {R.}~\bibnamefont
  {Lipinski~Jusinskas}}, \ and\ \bibinfo {author} {\bibfnamefont
  {A.}~\bibnamefont {Lipstein}},\ }\href {\doibase 10.1007/JHEP07(2022)004}
  {\bibfield  {journal} {\bibinfo  {journal} {JHEP}\ }\textbf {\bibinfo
  {volume} {07}},\ \bibinfo {pages} {004} (\bibinfo {year} {2022})},\ \Eprint
  {http://arxiv.org/abs/2112.12695} {arXiv:2112.12695 [hep-th]} \BibitemShut
  {NoStop}%
\bibitem [{\citenamefont {Herderschee}(2021)}]{Herderschee:2021jbi}%
  \BibitemOpen
  \bibfield  {author} {\bibinfo {author} {\bibfnamefont {A.}~\bibnamefont
  {Herderschee}},\ }\href@noop {} {\  (\bibinfo {year} {2021})},\ \Eprint
  {http://arxiv.org/abs/2112.08226} {arXiv:2112.08226 [hep-th]} \BibitemShut
  {NoStop}%
\bibitem [{\citenamefont {Maldacena}(2003)}]{Maldacena:2002vr}%
  \BibitemOpen
  \bibfield  {author} {\bibinfo {author} {\bibfnamefont {J.~M.}\ \bibnamefont
  {Maldacena}},\ }\href {\doibase 10.1088/1126-6708/2003/05/013} {\bibfield
  {journal} {\bibinfo  {journal} {JHEP}\ }\textbf {\bibinfo {volume} {05}},\
  \bibinfo {pages} {013} (\bibinfo {year} {2003})},\ \Eprint
  {http://arxiv.org/abs/astro-ph/0210603} {arXiv:astro-ph/0210603} \BibitemShut
  {NoStop}%
\bibitem [{\citenamefont {Lee}\ and\ \citenamefont {Wang}(2022)}]{Lee:2022fgr}%
  \BibitemOpen
  \bibfield  {author} {\bibinfo {author} {\bibfnamefont {H.}~\bibnamefont
  {Lee}}\ and\ \bibinfo {author} {\bibfnamefont {X.}~\bibnamefont {Wang}},\
  }\href@noop {} {\  (\bibinfo {year} {2022})},\ \Eprint
  {http://arxiv.org/abs/2212.11282} {arXiv:2212.11282 [hep-th]} \BibitemShut
  {NoStop}%
\bibitem [{\citenamefont {Karateev}\ \emph {et~al.}(2018)\citenamefont
  {Karateev}, \citenamefont {Kravchuk},\ and\ \citenamefont
  {Simmons-Duffin}}]{Karateev:2017jgd}%
  \BibitemOpen
  \bibfield  {author} {\bibinfo {author} {\bibfnamefont {D.}~\bibnamefont
  {Karateev}}, \bibinfo {author} {\bibfnamefont {P.}~\bibnamefont {Kravchuk}},
  \ and\ \bibinfo {author} {\bibfnamefont {D.}~\bibnamefont {Simmons-Duffin}},\
  }\href {\doibase 10.1007/JHEP02(2018)081} {\bibfield  {journal} {\bibinfo
  {journal} {JHEP}\ }\textbf {\bibinfo {volume} {02}},\ \bibinfo {pages} {081}
  (\bibinfo {year} {2018})},\ \Eprint {http://arxiv.org/abs/1706.07813}
  {arXiv:1706.07813 [hep-th]} \BibitemShut {NoStop}%
\bibitem [{\citenamefont {Costa}\ \emph {et~al.}(2011)\citenamefont {Costa},
  \citenamefont {Penedones}, \citenamefont {Poland},\ and\ \citenamefont
  {Rychkov}}]{Costa:2011mg}%
  \BibitemOpen
  \bibfield  {author} {\bibinfo {author} {\bibfnamefont {M.~S.}\ \bibnamefont
  {Costa}}, \bibinfo {author} {\bibfnamefont {J.}~\bibnamefont {Penedones}},
  \bibinfo {author} {\bibfnamefont {D.}~\bibnamefont {Poland}}, \ and\ \bibinfo
  {author} {\bibfnamefont {S.}~\bibnamefont {Rychkov}},\ }\href {\doibase
  10.1007/JHEP11(2011)071} {\bibfield  {journal} {\bibinfo  {journal} {JHEP}\
  }\textbf {\bibinfo {volume} {11}},\ \bibinfo {pages} {071} (\bibinfo {year}
  {2011})},\ \Eprint {http://arxiv.org/abs/1107.3554} {arXiv:1107.3554
  [hep-th]} \BibitemShut {NoStop}%
\bibitem [{\citenamefont {Kravchuk}\ and\ \citenamefont
  {Simmons-Duffin}(2018{\natexlab{a}})}]{Kravchuk:2016qvl}%
  \BibitemOpen
  \bibfield  {author} {\bibinfo {author} {\bibfnamefont {P.}~\bibnamefont
  {Kravchuk}}\ and\ \bibinfo {author} {\bibfnamefont {D.}~\bibnamefont
  {Simmons-Duffin}},\ }\href {\doibase 10.1007/JHEP02(2018)096} {\bibfield
  {journal} {\bibinfo  {journal} {JHEP}\ }\textbf {\bibinfo {volume} {02}},\
  \bibinfo {pages} {096} (\bibinfo {year} {2018}{\natexlab{a}})},\ \Eprint
  {http://arxiv.org/abs/1612.08987} {arXiv:1612.08987 [hep-th]} \BibitemShut
  {NoStop}%
\bibitem [{\citenamefont {Karateev}\ \emph {et~al.}(2019)\citenamefont
  {Karateev}, \citenamefont {Kravchuk},\ and\ \citenamefont
  {Simmons-Duffin}}]{Karateev:2018oml}%
  \BibitemOpen
  \bibfield  {author} {\bibinfo {author} {\bibfnamefont {D.}~\bibnamefont
  {Karateev}}, \bibinfo {author} {\bibfnamefont {P.}~\bibnamefont {Kravchuk}},
  \ and\ \bibinfo {author} {\bibfnamefont {D.}~\bibnamefont {Simmons-Duffin}},\
  }\href {\doibase 10.1007/JHEP10(2019)217} {\bibfield  {journal} {\bibinfo
  {journal} {JHEP}\ }\textbf {\bibinfo {volume} {10}},\ \bibinfo {pages} {217}
  (\bibinfo {year} {2019})},\ \Eprint {http://arxiv.org/abs/1809.05111}
  {arXiv:1809.05111 [hep-th]} \BibitemShut {NoStop}%
\bibitem [{\citenamefont {Caron-Huot}\ and\ \citenamefont
  {Li}(2021)}]{Caron-Huot:2021kjy}%
  \BibitemOpen
  \bibfield  {author} {\bibinfo {author} {\bibfnamefont {S.}~\bibnamefont
  {Caron-Huot}}\ and\ \bibinfo {author} {\bibfnamefont {Y.-Z.}\ \bibnamefont
  {Li}},\ }\href {\doibase 10.1007/JHEP06(2021)041} {\bibfield  {journal}
  {\bibinfo  {journal} {JHEP}\ }\textbf {\bibinfo {volume} {06}},\ \bibinfo
  {pages} {041} (\bibinfo {year} {2021})},\ \Eprint
  {http://arxiv.org/abs/2102.08160} {arXiv:2102.08160 [hep-th]} \BibitemShut
  {NoStop}%
\bibitem [{\citenamefont {Witten}(1998)}]{Witten:1998qj}%
  \BibitemOpen
  \bibfield  {author} {\bibinfo {author} {\bibfnamefont {E.}~\bibnamefont
  {Witten}},\ }\href {\doibase 10.4310/ATMP.1998.v2.n2.a2} {\bibfield
  {journal} {\bibinfo  {journal} {Adv. Theor. Math. Phys.}\ }\textbf {\bibinfo
  {volume} {2}},\ \bibinfo {pages} {253} (\bibinfo {year} {1998})},\ \Eprint
  {http://arxiv.org/abs/hep-th/9802150} {arXiv:hep-th/9802150} \BibitemShut
  {NoStop}%
\bibitem [{\citenamefont {Gubser}\ \emph {et~al.}(1998)\citenamefont {Gubser},
  \citenamefont {Klebanov},\ and\ \citenamefont {Polyakov}}]{Gubser:1998bc}%
  \BibitemOpen
  \bibfield  {author} {\bibinfo {author} {\bibfnamefont {S.~S.}\ \bibnamefont
  {Gubser}}, \bibinfo {author} {\bibfnamefont {I.~R.}\ \bibnamefont
  {Klebanov}}, \ and\ \bibinfo {author} {\bibfnamefont {A.~M.}\ \bibnamefont
  {Polyakov}},\ }\href {\doibase 10.1016/S0370-2693(98)00377-3} {\bibfield
  {journal} {\bibinfo  {journal} {Phys. Lett. B}\ }\textbf {\bibinfo {volume}
  {428}},\ \bibinfo {pages} {105} (\bibinfo {year} {1998})},\ \Eprint
  {http://arxiv.org/abs/hep-th/9802109} {arXiv:hep-th/9802109} \BibitemShut
  {NoStop}%
\bibitem [{\citenamefont {Costa}\ \emph {et~al.}(2014)\citenamefont {Costa},
  \citenamefont {Goncalves},\ and\ \citenamefont {Penedones}}]{Costa:2014kfa}%
  \BibitemOpen
  \bibfield  {author} {\bibinfo {author} {\bibfnamefont {M.~S.}\ \bibnamefont
  {Costa}}, \bibinfo {author} {\bibfnamefont {V.}~\bibnamefont {Goncalves}}, \
  and\ \bibinfo {author} {\bibfnamefont {J.}~\bibnamefont {Penedones}},\ }\href
  {\doibase 10.1007/JHEP09(2014)064} {\bibfield  {journal} {\bibinfo  {journal}
  {JHEP}\ }\textbf {\bibinfo {volume} {09}},\ \bibinfo {pages} {064} (\bibinfo
  {year} {2014})},\ \Eprint {http://arxiv.org/abs/1404.5625} {arXiv:1404.5625
  [hep-th]} \BibitemShut {NoStop}%
\bibitem [{\citenamefont {Raju}(2012{\natexlab{b}})}]{Raju:2012zs}%
  \BibitemOpen
  \bibfield  {author} {\bibinfo {author} {\bibfnamefont {S.}~\bibnamefont
  {Raju}},\ }\href {\doibase 10.1103/PhysRevD.85.126008} {\bibfield  {journal}
  {\bibinfo  {journal} {Phys. Rev. D}\ }\textbf {\bibinfo {volume} {85}},\
  \bibinfo {pages} {126008} (\bibinfo {year} {2012}{\natexlab{b}})},\ \Eprint
  {http://arxiv.org/abs/1201.6452} {arXiv:1201.6452 [hep-th]} \BibitemShut
  {NoStop}%
\bibitem [{\citenamefont {Li}\ and\ \citenamefont {Mei}()}]{YueZhou:future}%
  \BibitemOpen
  \bibfield  {author} {\bibinfo {author} {\bibfnamefont {Y.-Z.}\ \bibnamefont
  {Li}}\ and\ \bibinfo {author} {\bibfnamefont {J.}~\bibnamefont {Mei}},\
  }\href@noop {} {\bibinfo  {journal} {to appear}\ }\BibitemShut {NoStop}%
\bibitem [{Note3()}]{Note3}%
  \BibitemOpen
\bibfield  {journal} {  }\bibinfo {note} {These identities can trivially pass
  the strange operator $1/\protect \mathcal {D}_{ij}$, because the Ward
  identity always allows us to change $(ij)$ legs to other legs that $\partial
  _X\cdot \protect \mathcal {D}_Z$ does not act.}\BibitemShut {Stop}%
\bibitem [{Note4()}]{Note4}%
  \BibitemOpen
  \bibinfo {note} {Since the replacement $\protect \mathcal {E}\rightarrow
  \protect \mathcal {P}$ makes even YM amplitude not linear in $\protect
  \mathcal {P}$, the simple representation using conformal generator proposed
  in sec \ref {sec: BCJ} does not work.}\BibitemShut {Stop}%
\bibitem [{\citenamefont {Simmons-Duffin}(2014)}]{SimmonsDuffin:2012uy}%
  \BibitemOpen
  \bibfield  {author} {\bibinfo {author} {\bibfnamefont {D.}~\bibnamefont
  {Simmons-Duffin}},\ }\href {\doibase 10.1007/JHEP04(2014)146} {\bibfield
  {journal} {\bibinfo  {journal} {JHEP}\ }\textbf {\bibinfo {volume} {04}},\
  \bibinfo {pages} {146} (\bibinfo {year} {2014})},\ \Eprint
  {http://arxiv.org/abs/1204.3894} {arXiv:1204.3894 [hep-th]} \BibitemShut
  {NoStop}%
\bibitem [{\citenamefont {Li}\ and\ \citenamefont {Mei}(2023)}]{Li:2023azu}%
  \BibitemOpen
  \bibfield  {author} {\bibinfo {author} {\bibfnamefont {Y.-Z.}\ \bibnamefont
  {Li}}\ and\ \bibinfo {author} {\bibfnamefont {J.}~\bibnamefont {Mei}},\
  }\href@noop {} {\  (\bibinfo {year} {2023})},\ \Eprint
  {http://arxiv.org/abs/2304.12757} {arXiv:2304.12757 [hep-th]} \BibitemShut
  {NoStop}%
\bibitem [{\citenamefont {Mack}(2009)}]{Mack:2009mi}%
  \BibitemOpen
  \bibfield  {author} {\bibinfo {author} {\bibfnamefont {G.}~\bibnamefont
  {Mack}},\ }\href@noop {} {\  (\bibinfo {year} {2009})},\ \Eprint
  {http://arxiv.org/abs/0907.2407} {arXiv:0907.2407 [hep-th]} \BibitemShut
  {NoStop}%
\bibitem [{\citenamefont {Armstrong}\ \emph {et~al.}(2021)\citenamefont
  {Armstrong}, \citenamefont {Lipstein},\ and\ \citenamefont
  {Mei}}]{Armstrong:2020woi}%
  \BibitemOpen
  \bibfield  {author} {\bibinfo {author} {\bibfnamefont {C.}~\bibnamefont
  {Armstrong}}, \bibinfo {author} {\bibfnamefont {A.~E.}\ \bibnamefont
  {Lipstein}}, \ and\ \bibinfo {author} {\bibfnamefont {J.}~\bibnamefont
  {Mei}},\ }\href {\doibase 10.1007/JHEP02(2021)194} {\bibfield  {journal}
  {\bibinfo  {journal} {JHEP}\ }\textbf {\bibinfo {volume} {02}},\ \bibinfo
  {pages} {194} (\bibinfo {year} {2021})},\ \Eprint
  {http://arxiv.org/abs/2012.02059} {arXiv:2012.02059 [hep-th]} \BibitemShut
  {NoStop}%
\bibitem [{\citenamefont {Albayrak}\ \emph {et~al.}(2021)\citenamefont
  {Albayrak}, \citenamefont {Kharel},\ and\ \citenamefont
  {Meltzer}}]{Albayrak:2020fyp}%
  \BibitemOpen
  \bibfield  {author} {\bibinfo {author} {\bibfnamefont {S.}~\bibnamefont
  {Albayrak}}, \bibinfo {author} {\bibfnamefont {S.}~\bibnamefont {Kharel}}, \
  and\ \bibinfo {author} {\bibfnamefont {D.}~\bibnamefont {Meltzer}},\ }\href
  {\doibase 10.1007/JHEP03(2021)249} {\bibfield  {journal} {\bibinfo  {journal}
  {JHEP}\ }\textbf {\bibinfo {volume} {03}},\ \bibinfo {pages} {249} (\bibinfo
  {year} {2021})},\ \Eprint {http://arxiv.org/abs/2012.10460} {arXiv:2012.10460
  [hep-th]} \BibitemShut {NoStop}%
\bibitem [{\citenamefont {Farrow}\ \emph {et~al.}(2019)\citenamefont {Farrow},
  \citenamefont {Lipstein},\ and\ \citenamefont {McFadden}}]{Farrow:2018yni}%
  \BibitemOpen
  \bibfield  {author} {\bibinfo {author} {\bibfnamefont {J.~A.}\ \bibnamefont
  {Farrow}}, \bibinfo {author} {\bibfnamefont {A.~E.}\ \bibnamefont
  {Lipstein}}, \ and\ \bibinfo {author} {\bibfnamefont {P.}~\bibnamefont
  {McFadden}},\ }\href {\doibase 10.1007/JHEP02(2019)130} {\bibfield  {journal}
  {\bibinfo  {journal} {JHEP}\ }\textbf {\bibinfo {volume} {02}},\ \bibinfo
  {pages} {130} (\bibinfo {year} {2019})},\ \Eprint
  {http://arxiv.org/abs/1812.11129} {arXiv:1812.11129 [hep-th]} \BibitemShut
  {NoStop}%
\bibitem [{\citenamefont {Armstrong}\ \emph
  {et~al.}(2022{\natexlab{a}})\citenamefont {Armstrong}, \citenamefont {Gomez},
  \citenamefont {Lipinski~Jusinskas}, \citenamefont {Lipstein},\ and\
  \citenamefont {Mei}}]{Armstrong:2022mfr}%
  \BibitemOpen
  \bibfield  {author} {\bibinfo {author} {\bibfnamefont {C.}~\bibnamefont
  {Armstrong}}, \bibinfo {author} {\bibfnamefont {H.}~\bibnamefont {Gomez}},
  \bibinfo {author} {\bibfnamefont {R.}~\bibnamefont {Lipinski~Jusinskas}},
  \bibinfo {author} {\bibfnamefont {A.}~\bibnamefont {Lipstein}}, \ and\
  \bibinfo {author} {\bibfnamefont {J.}~\bibnamefont {Mei}},\ }\href {\doibase
  10.1103/PhysRevD.106.L121701} {\bibfield  {journal} {\bibinfo  {journal}
  {Phys. Rev. D}\ }\textbf {\bibinfo {volume} {106}},\ \bibinfo {pages}
  {L121701} (\bibinfo {year} {2022}{\natexlab{a}})},\ \Eprint
  {http://arxiv.org/abs/2209.02709} {arXiv:2209.02709 [hep-th]} \BibitemShut
  {NoStop}%
\bibitem [{\citenamefont {Armstrong}\ \emph
  {et~al.}(2022{\natexlab{b}})\citenamefont {Armstrong}, \citenamefont {Gomez},
  \citenamefont {Lipinski~Jusinskas}, \citenamefont {Lipstein},\ and\
  \citenamefont {Mei}}]{Armstrong:2022csc}%
  \BibitemOpen
  \bibfield  {author} {\bibinfo {author} {\bibfnamefont {C.}~\bibnamefont
  {Armstrong}}, \bibinfo {author} {\bibfnamefont {H.}~\bibnamefont {Gomez}},
  \bibinfo {author} {\bibfnamefont {R.}~\bibnamefont {Lipinski~Jusinskas}},
  \bibinfo {author} {\bibfnamefont {A.}~\bibnamefont {Lipstein}}, \ and\
  \bibinfo {author} {\bibfnamefont {J.}~\bibnamefont {Mei}},\ }\href {\doibase
  10.1007/JHEP08(2022)054} {\bibfield  {journal} {\bibinfo  {journal} {JHEP}\
  }\textbf {\bibinfo {volume} {08}},\ \bibinfo {pages} {054} (\bibinfo {year}
  {2022}{\natexlab{b}})},\ \Eprint {http://arxiv.org/abs/2204.08931}
  {arXiv:2204.08931 [hep-th]} \BibitemShut {NoStop}%
\bibitem [{\citenamefont {Alday}\ \emph {et~al.}(2021)\citenamefont {Alday},
  \citenamefont {Behan}, \citenamefont {Ferrero},\ and\ \citenamefont
  {Zhou}}]{Alday:2021odx}%
  \BibitemOpen
  \bibfield  {author} {\bibinfo {author} {\bibfnamefont {L.~F.}\ \bibnamefont
  {Alday}}, \bibinfo {author} {\bibfnamefont {C.}~\bibnamefont {Behan}},
  \bibinfo {author} {\bibfnamefont {P.}~\bibnamefont {Ferrero}}, \ and\
  \bibinfo {author} {\bibfnamefont {X.}~\bibnamefont {Zhou}},\ }\href {\doibase
  10.1007/JHEP06(2021)020} {\bibfield  {journal} {\bibinfo  {journal} {JHEP}\
  }\textbf {\bibinfo {volume} {06}},\ \bibinfo {pages} {020} (\bibinfo {year}
  {2021})},\ \Eprint {http://arxiv.org/abs/2103.15830} {arXiv:2103.15830
  [hep-th]} \BibitemShut {NoStop}%
\bibitem [{\citenamefont {Zhou}(2021)}]{Zhou:2021gnu}%
  \BibitemOpen
  \bibfield  {author} {\bibinfo {author} {\bibfnamefont {X.}~\bibnamefont
  {Zhou}},\ }\href {\doibase 10.1103/PhysRevLett.127.141601} {\bibfield
  {journal} {\bibinfo  {journal} {Phys. Rev. Lett.}\ }\textbf {\bibinfo
  {volume} {127}},\ \bibinfo {pages} {141601} (\bibinfo {year} {2021})},\
  \Eprint {http://arxiv.org/abs/2106.07651} {arXiv:2106.07651 [hep-th]}
  \BibitemShut {NoStop}%
\bibitem [{\citenamefont {Bissi}\ \emph
  {et~al.}(2022{\natexlab{b}})\citenamefont {Bissi}, \citenamefont {Fardelli},
  \citenamefont {Manenti},\ and\ \citenamefont {Zhou}}]{Bissi:2022wuh}%
  \BibitemOpen
  \bibfield  {author} {\bibinfo {author} {\bibfnamefont {A.}~\bibnamefont
  {Bissi}}, \bibinfo {author} {\bibfnamefont {G.}~\bibnamefont {Fardelli}},
  \bibinfo {author} {\bibfnamefont {A.}~\bibnamefont {Manenti}}, \ and\
  \bibinfo {author} {\bibfnamefont {X.}~\bibnamefont {Zhou}},\ }\href@noop {}
  {\  (\bibinfo {year} {2022}{\natexlab{b}})},\ \Eprint
  {http://arxiv.org/abs/2209.01204} {arXiv:2209.01204 [hep-th]} \BibitemShut
  {NoStop}%
\bibitem [{\citenamefont {Sleight}\ and\ \citenamefont
  {Taronna}(2021{\natexlab{a}})}]{Sleight:2020obc}%
  \BibitemOpen
  \bibfield  {author} {\bibinfo {author} {\bibfnamefont {C.}~\bibnamefont
  {Sleight}}\ and\ \bibinfo {author} {\bibfnamefont {M.}~\bibnamefont
  {Taronna}},\ }\href {\doibase 10.1103/PhysRevD.104.L081902} {\bibfield
  {journal} {\bibinfo  {journal} {Phys. Rev. D}\ }\textbf {\bibinfo {volume}
  {104}},\ \bibinfo {pages} {L081902} (\bibinfo {year} {2021}{\natexlab{a}})},\
  \Eprint {http://arxiv.org/abs/2007.09993} {arXiv:2007.09993 [hep-th]}
  \BibitemShut {NoStop}%
\bibitem [{\citenamefont {Sleight}\ and\ \citenamefont
  {Taronna}(2021{\natexlab{b}})}]{Sleight:2021plv}%
  \BibitemOpen
  \bibfield  {author} {\bibinfo {author} {\bibfnamefont {C.}~\bibnamefont
  {Sleight}}\ and\ \bibinfo {author} {\bibfnamefont {M.}~\bibnamefont
  {Taronna}},\ }\href {\doibase 10.1007/JHEP12(2021)074} {\bibfield  {journal}
  {\bibinfo  {journal} {JHEP}\ }\textbf {\bibinfo {volume} {12}},\ \bibinfo
  {pages} {074} (\bibinfo {year} {2021}{\natexlab{b}})},\ \Eprint
  {http://arxiv.org/abs/2109.02725} {arXiv:2109.02725 [hep-th]} \BibitemShut
  {NoStop}%
\bibitem [{\citenamefont {Baumann}\ \emph {et~al.}(2022)\citenamefont
  {Baumann}, \citenamefont {Green}, \citenamefont {Joyce}, \citenamefont
  {Pajer}, \citenamefont {Pimentel}, \citenamefont {Sleight},\ and\
  \citenamefont {Taronna}}]{Baumann:2022jpr}%
  \BibitemOpen
  \bibfield  {author} {\bibinfo {author} {\bibfnamefont {D.}~\bibnamefont
  {Baumann}}, \bibinfo {author} {\bibfnamefont {D.}~\bibnamefont {Green}},
  \bibinfo {author} {\bibfnamefont {A.}~\bibnamefont {Joyce}}, \bibinfo
  {author} {\bibfnamefont {E.}~\bibnamefont {Pajer}}, \bibinfo {author}
  {\bibfnamefont {G.~L.}\ \bibnamefont {Pimentel}}, \bibinfo {author}
  {\bibfnamefont {C.}~\bibnamefont {Sleight}}, \ and\ \bibinfo {author}
  {\bibfnamefont {M.}~\bibnamefont {Taronna}},\ }in\ \href@noop {} {\emph
  {\bibinfo {booktitle} {{2022 Snowmass Summer Study}}}}\ (\bibinfo {year}
  {2022})\ \Eprint {http://arxiv.org/abs/2203.08121} {arXiv:2203.08121
  [hep-th]} \BibitemShut {NoStop}%
\bibitem [{\citenamefont {Sleight}(2020)}]{Sleight:2019mgd}%
  \BibitemOpen
  \bibfield  {author} {\bibinfo {author} {\bibfnamefont {C.}~\bibnamefont
  {Sleight}},\ }\href {\doibase 10.1007/JHEP01(2020)090} {\bibfield  {journal}
  {\bibinfo  {journal} {JHEP}\ }\textbf {\bibinfo {volume} {01}},\ \bibinfo
  {pages} {090} (\bibinfo {year} {2020})},\ \Eprint
  {http://arxiv.org/abs/1906.12302} {arXiv:1906.12302 [hep-th]} \BibitemShut
  {NoStop}%
\bibitem [{\citenamefont {Arkani-Hamed}\ \emph {et~al.}(2020)\citenamefont
  {Arkani-Hamed}, \citenamefont {Baumann}, \citenamefont {Lee},\ and\
  \citenamefont {Pimentel}}]{Arkani-Hamed:2018kmz}%
  \BibitemOpen
  \bibfield  {author} {\bibinfo {author} {\bibfnamefont {N.}~\bibnamefont
  {Arkani-Hamed}}, \bibinfo {author} {\bibfnamefont {D.}~\bibnamefont
  {Baumann}}, \bibinfo {author} {\bibfnamefont {H.}~\bibnamefont {Lee}}, \ and\
  \bibinfo {author} {\bibfnamefont {G.~L.}\ \bibnamefont {Pimentel}},\ }\href
  {\doibase 10.1007/JHEP04(2020)105} {\bibfield  {journal} {\bibinfo  {journal}
  {JHEP}\ }\textbf {\bibinfo {volume} {04}},\ \bibinfo {pages} {105} (\bibinfo
  {year} {2020})},\ \Eprint {http://arxiv.org/abs/1811.00024} {arXiv:1811.00024
  [hep-th]} \BibitemShut {NoStop}%
\bibitem [{\citenamefont {Baumann}\ \emph
  {et~al.}(2020{\natexlab{a}})\citenamefont {Baumann}, \citenamefont
  {Duaso~Pueyo}, \citenamefont {Joyce}, \citenamefont {Lee},\ and\
  \citenamefont {Pimentel}}]{Baumann:2019oyu}%
  \BibitemOpen
  \bibfield  {author} {\bibinfo {author} {\bibfnamefont {D.}~\bibnamefont
  {Baumann}}, \bibinfo {author} {\bibfnamefont {C.}~\bibnamefont
  {Duaso~Pueyo}}, \bibinfo {author} {\bibfnamefont {A.}~\bibnamefont {Joyce}},
  \bibinfo {author} {\bibfnamefont {H.}~\bibnamefont {Lee}}, \ and\ \bibinfo
  {author} {\bibfnamefont {G.~L.}\ \bibnamefont {Pimentel}},\ }\href {\doibase
  10.1007/JHEP12(2020)204} {\bibfield  {journal} {\bibinfo  {journal} {JHEP}\
  }\textbf {\bibinfo {volume} {12}},\ \bibinfo {pages} {204} (\bibinfo {year}
  {2020}{\natexlab{a}})},\ \Eprint {http://arxiv.org/abs/1910.14051}
  {arXiv:1910.14051 [hep-th]} \BibitemShut {NoStop}%
\bibitem [{\citenamefont {Baumann}\ \emph
  {et~al.}(2020{\natexlab{b}})\citenamefont {Baumann}, \citenamefont
  {Duaso~Pueyo},\ and\ \citenamefont {Joyce}}]{Baumann:2020ksv}%
  \BibitemOpen
  \bibfield  {author} {\bibinfo {author} {\bibfnamefont {D.}~\bibnamefont
  {Baumann}}, \bibinfo {author} {\bibfnamefont {C.}~\bibnamefont
  {Duaso~Pueyo}}, \ and\ \bibinfo {author} {\bibfnamefont {A.}~\bibnamefont
  {Joyce}},\ }\href {\doibase 10.22661/AAPPSBL.2020.30.6.02} {\bibfield
  {journal} {\bibinfo  {journal} {AAPPS Bull.}\ }\textbf {\bibinfo {volume}
  {30}},\ \bibinfo {pages} {2} (\bibinfo {year}
  {2020}{\natexlab{b}})}\BibitemShut {NoStop}%
\bibitem [{\citenamefont {Baumann}\ \emph {et~al.}(2021)\citenamefont
  {Baumann}, \citenamefont {Duaso~Pueyo}, \citenamefont {Joyce}, \citenamefont
  {Lee},\ and\ \citenamefont {Pimentel}}]{Baumann:2020dch}%
  \BibitemOpen
  \bibfield  {author} {\bibinfo {author} {\bibfnamefont {D.}~\bibnamefont
  {Baumann}}, \bibinfo {author} {\bibfnamefont {C.}~\bibnamefont
  {Duaso~Pueyo}}, \bibinfo {author} {\bibfnamefont {A.}~\bibnamefont {Joyce}},
  \bibinfo {author} {\bibfnamefont {H.}~\bibnamefont {Lee}}, \ and\ \bibinfo
  {author} {\bibfnamefont {G.~L.}\ \bibnamefont {Pimentel}},\ }\href {\doibase
  10.21468/SciPostPhys.11.3.071} {\bibfield  {journal} {\bibinfo  {journal}
  {SciPost Phys.}\ }\textbf {\bibinfo {volume} {11}},\ \bibinfo {pages} {071}
  (\bibinfo {year} {2021})},\ \Eprint {http://arxiv.org/abs/2005.04234}
  {arXiv:2005.04234 [hep-th]} \BibitemShut {NoStop}%
\bibitem [{\citenamefont {Binder}\ \emph {et~al.}(2022)\citenamefont {Binder},
  \citenamefont {Freedman},\ and\ \citenamefont {Pufu}}]{Binder:2020raz}%
  \BibitemOpen
  \bibfield  {author} {\bibinfo {author} {\bibfnamefont {D.~J.}\ \bibnamefont
  {Binder}}, \bibinfo {author} {\bibfnamefont {D.~Z.}\ \bibnamefont
  {Freedman}}, \ and\ \bibinfo {author} {\bibfnamefont {S.~S.}\ \bibnamefont
  {Pufu}},\ }\href {\doibase 10.1007/JHEP02(2022)040} {\bibfield  {journal}
  {\bibinfo  {journal} {JHEP}\ }\textbf {\bibinfo {volume} {02}},\ \bibinfo
  {pages} {040} (\bibinfo {year} {2022})},\ \Eprint
  {http://arxiv.org/abs/2003.07448} {arXiv:2003.07448 [hep-th]} \BibitemShut
  {NoStop}%
\bibitem [{\citenamefont {Goncalves}\ \emph {et~al.}(2015)\citenamefont
  {Goncalves}, \citenamefont {Penedones},\ and\ \citenamefont
  {Trevisani}}]{Goncalves:2014rfa}%
  \BibitemOpen
  \bibfield  {author} {\bibinfo {author} {\bibfnamefont {V.}~\bibnamefont
  {Goncalves}}, \bibinfo {author} {\bibfnamefont {J.}~\bibnamefont
  {Penedones}}, \ and\ \bibinfo {author} {\bibfnamefont {E.}~\bibnamefont
  {Trevisani}},\ }\href {\doibase 10.1007/JHEP10(2015)040} {\bibfield
  {journal} {\bibinfo  {journal} {JHEP}\ }\textbf {\bibinfo {volume} {10}},\
  \bibinfo {pages} {040} (\bibinfo {year} {2015})},\ \Eprint
  {http://arxiv.org/abs/1410.4185} {arXiv:1410.4185 [hep-th]} \BibitemShut
  {NoStop}%
\bibitem [{\citenamefont {van Rees}\ and\ \citenamefont
  {Zhao}(2022)}]{vanRees:2022itk}%
  \BibitemOpen
  \bibfield  {author} {\bibinfo {author} {\bibfnamefont {B.~C.}\ \bibnamefont
  {van Rees}}\ and\ \bibinfo {author} {\bibfnamefont {X.}~\bibnamefont
  {Zhao}},\ }\href@noop {} {\  (\bibinfo {year} {2022})},\ \Eprint
  {http://arxiv.org/abs/2210.15683} {arXiv:2210.15683 [hep-th]} \BibitemShut
  {NoStop}%
\bibitem [{\citenamefont {Caron-Huot}\ \emph {et~al.}(2021)\citenamefont
  {Caron-Huot}, \citenamefont {Mazac}, \citenamefont {Rastelli},\ and\
  \citenamefont {Simmons-Duffin}}]{Caron-Huot:2021enk}%
  \BibitemOpen
  \bibfield  {author} {\bibinfo {author} {\bibfnamefont {S.}~\bibnamefont
  {Caron-Huot}}, \bibinfo {author} {\bibfnamefont {D.}~\bibnamefont {Mazac}},
  \bibinfo {author} {\bibfnamefont {L.}~\bibnamefont {Rastelli}}, \ and\
  \bibinfo {author} {\bibfnamefont {D.}~\bibnamefont {Simmons-Duffin}},\ }\href
  {\doibase 10.1007/JHEP11(2021)164} {\bibfield  {journal} {\bibinfo  {journal}
  {JHEP}\ }\textbf {\bibinfo {volume} {11}},\ \bibinfo {pages} {164} (\bibinfo
  {year} {2021})},\ \Eprint {http://arxiv.org/abs/2106.10274} {arXiv:2106.10274
  [hep-th]} \BibitemShut {NoStop}%
\bibitem [{\citenamefont {Duary}\ \emph {et~al.}(2022)\citenamefont {Duary},
  \citenamefont {Hijano},\ and\ \citenamefont {Patra}}]{Duary:2022pyv}%
  \BibitemOpen
  \bibfield  {author} {\bibinfo {author} {\bibfnamefont {S.}~\bibnamefont
  {Duary}}, \bibinfo {author} {\bibfnamefont {E.}~\bibnamefont {Hijano}}, \
  and\ \bibinfo {author} {\bibfnamefont {M.}~\bibnamefont {Patra}},\
  }\href@noop {} {\  (\bibinfo {year} {2022})},\ \Eprint
  {http://arxiv.org/abs/2211.13711} {arXiv:2211.13711 [hep-th]} \BibitemShut
  {NoStop}%
\bibitem [{\citenamefont {Duary}(2022)}]{Duary:2022afn}%
  \BibitemOpen
  \bibfield  {author} {\bibinfo {author} {\bibfnamefont {S.}~\bibnamefont
  {Duary}},\ }\href@noop {} {\  (\bibinfo {year} {2022})},\ \Eprint
  {http://arxiv.org/abs/2212.09509} {arXiv:2212.09509 [hep-th]} \BibitemShut
  {NoStop}%
\bibitem [{\citenamefont {Caron-Huot}(2017)}]{Caron-Huot:2017vep}%
  \BibitemOpen
  \bibfield  {author} {\bibinfo {author} {\bibfnamefont {S.}~\bibnamefont
  {Caron-Huot}},\ }\href {\doibase 10.1007/JHEP09(2017)078} {\bibfield
  {journal} {\bibinfo  {journal} {JHEP}\ }\textbf {\bibinfo {volume} {09}},\
  \bibinfo {pages} {078} (\bibinfo {year} {2017})},\ \Eprint
  {http://arxiv.org/abs/1703.00278} {arXiv:1703.00278 [hep-th]} \BibitemShut
  {NoStop}%
\bibitem [{\citenamefont {Simmons-Duffin}\ \emph {et~al.}(2018)\citenamefont
  {Simmons-Duffin}, \citenamefont {Stanford},\ and\ \citenamefont
  {Witten}}]{Simmons-Duffin:2017nub}%
  \BibitemOpen
  \bibfield  {author} {\bibinfo {author} {\bibfnamefont {D.}~\bibnamefont
  {Simmons-Duffin}}, \bibinfo {author} {\bibfnamefont {D.}~\bibnamefont
  {Stanford}}, \ and\ \bibinfo {author} {\bibfnamefont {E.}~\bibnamefont
  {Witten}},\ }\href {\doibase 10.1007/JHEP07(2018)085} {\bibfield  {journal}
  {\bibinfo  {journal} {JHEP}\ }\textbf {\bibinfo {volume} {07}},\ \bibinfo
  {pages} {085} (\bibinfo {year} {2018})},\ \Eprint
  {http://arxiv.org/abs/1711.03816} {arXiv:1711.03816 [hep-th]} \BibitemShut
  {NoStop}%
\bibitem [{\citenamefont {Kravchuk}\ and\ \citenamefont
  {Simmons-Duffin}(2018{\natexlab{b}})}]{Kravchuk:2018htv}%
  \BibitemOpen
  \bibfield  {author} {\bibinfo {author} {\bibfnamefont {P.}~\bibnamefont
  {Kravchuk}}\ and\ \bibinfo {author} {\bibfnamefont {D.}~\bibnamefont
  {Simmons-Duffin}},\ }\href {\doibase 10.1007/JHEP11(2018)102} {\bibfield
  {journal} {\bibinfo  {journal} {JHEP}\ }\textbf {\bibinfo {volume} {11}},\
  \bibinfo {pages} {102} (\bibinfo {year} {2018}{\natexlab{b}})},\ \bibinfo
  {note} {[,236(2018)]},\ \Eprint {http://arxiv.org/abs/1805.00098}
  {arXiv:1805.00098 [hep-th]} \BibitemShut {NoStop}%
\bibitem [{\citenamefont {Hogervorst}\ \emph {et~al.}(2021)\citenamefont
  {Hogervorst}, \citenamefont {Penedones},\ and\ \citenamefont
  {Vaziri}}]{Hogervorst:2021uvp}%
  \BibitemOpen
  \bibfield  {author} {\bibinfo {author} {\bibfnamefont {M.}~\bibnamefont
  {Hogervorst}}, \bibinfo {author} {\bibfnamefont {J.~a.}\ \bibnamefont
  {Penedones}}, \ and\ \bibinfo {author} {\bibfnamefont {K.~S.}\ \bibnamefont
  {Vaziri}},\ }\href@noop {} {\  (\bibinfo {year} {2021})},\ \Eprint
  {http://arxiv.org/abs/2107.13871} {arXiv:2107.13871 [hep-th]} \BibitemShut
  {NoStop}%
\bibitem [{\citenamefont {Camanho}\ \emph {et~al.}(2016)\citenamefont
  {Camanho}, \citenamefont {Edelstein}, \citenamefont {Maldacena},\ and\
  \citenamefont {Zhiboedov}}]{Camanho:2014apa}%
  \BibitemOpen
  \bibfield  {author} {\bibinfo {author} {\bibfnamefont {X.~O.}\ \bibnamefont
  {Camanho}}, \bibinfo {author} {\bibfnamefont {J.~D.}\ \bibnamefont
  {Edelstein}}, \bibinfo {author} {\bibfnamefont {J.}~\bibnamefont
  {Maldacena}}, \ and\ \bibinfo {author} {\bibfnamefont {A.}~\bibnamefont
  {Zhiboedov}},\ }\href {\doibase 10.1007/JHEP02(2016)020} {\bibfield
  {journal} {\bibinfo  {journal} {JHEP}\ }\textbf {\bibinfo {volume} {02}},\
  \bibinfo {pages} {020} (\bibinfo {year} {2016})},\ \Eprint
  {http://arxiv.org/abs/1407.5597} {arXiv:1407.5597 [hep-th]} \BibitemShut
  {NoStop}%
\bibitem [{\citenamefont {Henriksson}\ \emph {et~al.}(2022)\citenamefont
  {Henriksson}, \citenamefont {McPeak}, \citenamefont {Russo},\ and\
  \citenamefont {Vichi}}]{Henriksson:2022oeu}%
  \BibitemOpen
  \bibfield  {author} {\bibinfo {author} {\bibfnamefont {J.}~\bibnamefont
  {Henriksson}}, \bibinfo {author} {\bibfnamefont {B.}~\bibnamefont {McPeak}},
  \bibinfo {author} {\bibfnamefont {F.}~\bibnamefont {Russo}}, \ and\ \bibinfo
  {author} {\bibfnamefont {A.}~\bibnamefont {Vichi}},\ }\href@noop {} {\
  (\bibinfo {year} {2022})},\ \Eprint {http://arxiv.org/abs/2203.08164}
  {arXiv:2203.08164 [hep-th]} \BibitemShut {NoStop}%
\bibitem [{\citenamefont {Martin-Garcia}()}]{xact}%
  \BibitemOpen
  \bibfield  {author} {\bibinfo {author} {\bibfnamefont {J.~M.}\ \bibnamefont
  {Martin-Garcia}},\ }\href@noop {} {\bibinfo  {journal} {http://xact.es/}\
  }\BibitemShut {NoStop}%
\end{thebibliography}%

\onecolumngrid

\clearpage

\begin{appendix}

\section{Power counting in conformal correlators}
\label{app: power counting}

We elaborate on the discussion in section \ref{sec: building blocks} in this appendix. We start by considering the three-point functions of conserved currents. There are, in general, two parity-even OPE structures. These two three-point structures can be accounted by AdS YM three-point vertex \eqref{eq: YM three-pt vert} and a higher derivative cubic term arises in AdS EFTs
\be
\mathcal{L}=-\fft{f^{abc}}{3g_{\rm YM}^3} g_H F_{\mu}\,^{\nu a}F_{\nu}\,^{\rho b}F_{\rho}\,^{\mu c}\,.\label{eq: cub vert}
\ee
Using the bulk-to-boundary propagator of the gluon \eqref{eq: bulk-to-boundary A}, we can explicitly evaluate the resulting three-point functions by performing the integrals over AdS
\be
& \mathcal{M}_{3,{\rm YM}}\propto \fft{(2 d-3) \left(H_{23} V_1+H_{13} V_2\right)+V_3 \left((2 d-3) H_{12}+3 (d-2) V_1 V_2\right)}{\left(-2 X_1\cdot X_2\right)^{d/2} \left(-2 X_1\cdot X_3\right)^{d/2} \left(-2 X_2\cdot X_3\right)^{d/2}}\,,\nn\\
& \mathcal{M}_{3,{\rm cub}} \propto \fft{V_3 \left((d+2) V_1 V_2+H_{12}\right)+H_{23} V_1+H_{13} V_2}{\left(-2 X_1\cdot X_2\right)^{d/2} \left(-2 X_1\cdot X_3\right)^{d/2} \left(-2 X_2\cdot X_3\right)^{d/2}}\,,\label{eq: VVVs}
\ee
where the definitions of $H_{ij}$ and $V_i$ are
\be
H_{ij}=2\big((X_i\cdot Z_j)(Z_j \cdot X_i)-(X_i\cdot X_j)(Z_i \cdot Z_j)\big)\,,\quad V_i:=V_{i,jk}=\fft{(X_i\cdot X_k)(Z_i\cdot X_j)-(X_i\cdot X_j)(Z_i\cdot X_k)}{X_j\cdot X_k}\,.
\label{Hij}
\ee
See \cite{Costa:2011mg} for more details of this convention. We do not specify the overall coefficients in \eqref{eq: VVVs}, as they can be absorbed into the OPE coefficients. In holographic CFTs, the OPE associated with $\mathcal{M}_{3,{\rm cub}}$ should be suppressed by a large gap $\Delta_{\rm gap}\gg 1$ \cite{Camanho:2014apa}. However, staring at the LHS of \eqref{eq: VVVs} cannot tell which AdS vertices are their origins, making the power counting of OPE coefficients in terms of $1/\Delta_{\rm gap}$ vague. This is in contrast to the flat-space EFT amplitudes, where it is easy to figure out that higher power of the momentum comes with higher suppression of the UV scale $1/M$. 

Now as the differential operators $\mathcal{P}$ and $\mathcal{E}$ allow to uplift the flat-space amplitudes, the issue of the power counting in CFT can be resolved. We can show that $\mathcal{M}_{3,{\rm cub}} $ can also be uplifted from flat-space three-point amplitudes produced by \eqref{eq: cub vert}
\be
\hat{\mathcal{M}_{3,{\rm cub}} } \propto \mathcal{T}_{\mathcal{O}}\Big(\hat{F}_{1\mu}\,^\nu \hat{F}_{2\nu}\,^\rho \hat{F}_{3\rho}\,^\mu\Big) + \text{perm}\,,\quad \hat{F}_{i\mu\nu} = \mathcal{E}_{i\nu}\mathcal{P}_{i\mu}-\mathcal{E}_{i\mu}\mathcal{P}_{i\nu}\,.
\ee
It is then apparent that more $\mathcal{P}$ come with higher orders of $1/\Delta_{\rm gap}$ in holographic CFTs. It is worth noting that this flat-space structure persists even beyond holographic CFTs because three-point structures (not including the OPE) are general objects as they are essentially fixed by conformal symmetry. This makes the helicity structures of CFT$_3$ \cite{Caron-Huot:2021kjy} (i.e., the orthogonality of three-point structures) manifest by uplifting flat-space amplitudes.

We expect a similar uplift to work for three-point functions where the third operator is non-conserved. Such uplifts may diagonalize OPE matrix of mean field theory (MFT) in general dimensions, as one did for CFT$_3$ \cite{Caron-Huot:2021kjy}. Using the bootstrap idea, the diagonal MFT OPE could pave the way for constraining spinning correlation functions (beyond holographic CFTs).

The same arguments apply to the power counting in four-point correlators. For example, the low-lying terms in the four-point function of conserved currents from contact diagrams follow precisely as flat-space amplitudes (for the sharp power counting in flat-space where the gravity is dynamical, see \cite{Henriksson:2022oeu})
\be
\hat{\mathcal{M}}_{4V}^{\rm ct} \propto \fft{c_1}{C_T\Delta_{\rm gap}^2}\Big[\mathcal{T}_{\mathcal{O}}\Big(\hat{F}_{1\mu\nu}\hat{F}_2^{\mu\nu}\hat{F}_{3\rho\sigma}\hat{F}_4^{\rho\sigma}\Big)+{\rm perm}\Big] + \fft{c_2}{C_T\Delta_{\rm gap}^2} \Big[\mathcal{T}_{\mathcal{O}}\Big(\hat{F}_{1\mu}\,^{\nu}\hat{F}_{2\nu}\,^\rho\hat{F}_{3\rho}\,^{\sigma}\hat{F}_{4\sigma}\,^\mu\Big)\Big] + \cdots\,,
\ee
where $C_T$ is the central charge appearing as the coefficient of the stress-tensor two-point function. For holographic CFTs, we have the hierarchy $C_T\gg \Delta_{\rm gap}^{d-1}\gg 1$.

Although we focus on spin-$1$ conserved current in this appendix, the same principle should apply to three and four-point stress tensor correlators.

\section{Graviton vertices in AdS}
\label{app: vertex}

In this appendix, we provide three and four-point vertices for Einstein gravity in AdS
\be
S=\fft{1}{16\pi G}\int d^{d+1}x\sqrt{g}(R-2\Lambda)\,,
\ee
where $\Lambda=-(d-1)(d-2)/(2R_{\rm AdS}^2)$. To compute vertices, we expand the metric around the AdS background
\be
g_{\mu\nu} = g_{\mu\nu}^{\rm AdS}+\sqrt{32\pi G}h_{\mu\nu}\,,
\ee
and then expand the action up to the fourth order. For the three-point vertex, we perform \eqref{eq: def AdS amp} for two gravitons to give $\delta_{1,2} h$ while leaving the third one ``off-shell''. We vary off the off-shell graviton to give the three-point vertex function; for the four-point vertex, which gives rise to four-point contact amplitude, we use \eqref{eq: def AdS amp} to end up with the final answer. As in the main text, when it does not confuse, we slip off the superscript ``AdS'' and remember $g_{\mu\nu}$ is the AdS metric. We performed the calculations using the xAcT Mathematica package \cite{xact}.

\subsection{Three-point vertex}

The external on-shell gravitons are computed by the bulk-to-boundary propagators, which satisfy
\be
\delta_i h_\mu^\mu=\nabla_\mu\delta_ih^{\mu\nu}=0\,,\quad \Big(\nabla^2+\fft{2}{R_{\rm AdS}^2}\Big) \delta_ih_{\mu\nu}=0\,,
\ee
Then we find the three-point vertex can be written by
\be
V_{h,12}^{\mu\nu}= \hat{V}_{h,12}^{\mu\nu} - \fft{1}{2} \hat{V}_{h,12} g^{\mu\nu}\,,
\ee
where $\hat{V}_{h,12} $ is the trace of $\hat{V}_{h,12}^{\mu\nu}$
\be
\hat{V}_{h,12}^{\mu\nu}=&\sqrt{8\pi G}\Big(-\nabla^{\nu}\delta_2 h^{\rho \sigma }\nabla^{\mu}\delta_1 h_{\rho\sigma}-\nabla^{\mu}\delta_2 h^{\rho \sigma } \nabla^{\nu}\delta_1 h_{\rho\sigma}-2 \delta_1 h_{\rho\sigma}\nabla^{\sigma}\nabla^{\rho}\delta_2 h^{\mu \nu }+2 \delta_1 h_{\rho\sigma}\nabla^{\sigma}\nabla^{\nu}\delta_2 h^{\mu \rho }+2 \delta_1 h_{\rho\sigma}\nabla^{\sigma}\nabla^{\mu}\delta_2 h^{\nu \rho }\nn\\
&-2 \delta_1 h_{\rho\sigma} \nabla^{\nu}\nabla^{\mu}\delta_2 h^{\rho \sigma }-2 h_{2,\rho \sigma }\nabla^{\sigma}\nabla^{\rho}\delta_1 h^{\mu \nu }+2 h_{2,\rho \sigma }\nabla^{\sigma}\nabla^{\nu}\delta_1 h^{\mu \rho }+2 h_{2,\rho \sigma }\nabla^{\sigma}\nabla^{\mu}\delta_1 h^{\nu \rho }-2 h_{2,\rho \sigma } \nabla^{\nu}\nabla^{\mu}\delta_1 h^{\rho \sigma }\nn\\
&-2 \nabla _{\sigma }\delta_1 h^{\mu }_{\rho }\nabla^{\sigma}\delta_2 h^{\nu \rho }+2 \nabla _{\sigma }\delta_1 h^{\mu }_{\rho } \nabla^{\rho}\delta_2 h^{\nu \sigma }-2\nabla^{\sigma}\delta_2 h^{\mu \rho } \nabla _{\sigma }\delta_1 h^{\nu }_{\rho }+2 \nabla^{\rho}\delta_2 h^{\mu \sigma } \nabla _{\sigma }\delta_1 h^{\nu }_{\rho}\Big)\,.
\ee

\subsection{Four-point vertex}

Four-point vertex can be calculated similarly. We present the four-point vertex with ordering $(1234)$ below
\be
& V_{1234,h}=4\pi G\Big[\delta_1 h_{\mu\rho}\Big(2 \delta_2 h^{\mu \rho } \nabla_{\nu}\delta_4 h^{\gamma \sigma } \nabla _{\gamma }\delta_3 h_{\nu\sigma}-3 \delta_2 h^{\mu \rho } \nabla_{\gamma}\delta_4 h^{\nu \sigma } \nabla _{\gamma }\delta_3 h_{\nu\sigma}-16 \delta_2 h^{\mu }_{\sigma } \nabla_{\gamma}\delta_4 h^{\nu \sigma } \nabla_{\rho}\delta_3 h_{\gamma\nu}\nn\\
& +12 \delta_2 h^{\mu }_{\sigma } \nabla_{\rho}\delta_4 h^{\gamma \nu } \nabla_{\sigma}\delta_3 h_{\gamma\nu}-4 \delta_2 h_{\nu\sigma}\nabla_{\gamma}\delta_4 h^{\mu \rho } \nabla _{\gamma }\delta_3 h^{\nu \sigma }+12 \delta_2 h_{\nu\sigma}\nabla _{\gamma }\delta_3 h^{\mu \sigma } \nabla_{\gamma}\delta_4 h^{\nu \rho }-16 \delta_2 h_{\nu\sigma}\nabla_{\rho}\delta_3 h^{\mu }_{\gamma } \nabla_{\nu}\delta_4 h^{\gamma \sigma }\nn\\
&+24 \delta_2 h_{\nu\sigma}\nabla_{\sigma}\delta_3 h^{\mu }_{\gamma } \nabla_{\nu}\delta_4 h^{\gamma \rho }-16 \delta_2 h_{\nu\sigma}\nabla_{\sigma}\delta_3 h^{\mu }_{\gamma } \nabla_{\gamma}\delta_4 h^{\nu \rho }-8 \delta_2 h_{\nu\sigma}\nabla^{\mu }\delta_3 h^{\sigma }_{\gamma } \nabla_{\nu}\delta_4 h^{\gamma \rho }+16 \delta_2 h_{\nu\sigma}\nabla_{\gamma}\delta_4 h^{\mu \rho } \nabla_{\nu}\delta_3 h^{\sigma }_{\gamma }\nn\\&
+16 \delta_2 h^{\mu }_{\sigma } \delta_3 h_{\gamma\nu} \nabla_{\sigma}\nabla_{\rho}\delta_4 h^{\gamma \nu }-32 \delta_2 h^{\mu }_{\sigma } \delta_3 h_{\gamma\nu} \nabla_{\gamma}\nabla_{\rho}\delta_4 h^{\nu \sigma }+16 \delta_2 h^{\mu }_{\sigma } \delta_3 h_{\gamma\nu} \nabla_{\gamma}\nabla_{\nu}\delta_4 h^{\rho \sigma }-8 \delta_2 h^{\mu }_{\sigma } \nabla _{\gamma }\delta_3 h^{\rho }_{\nu } \nabla_{\nu}\delta_4 h^{\gamma \sigma }\nn\\
&+24 \delta_2 h^{\mu }_{\sigma } \nabla _{\gamma }\delta_3 h^{\rho }_{\nu } \nabla_{\gamma}\delta_4 h^{\nu \sigma }\Big)+\fft{\delta_1 h_{\mu\rho}}{R_{\rm AdS}^2}\Big(-d \delta_2 h^{\mu \rho } \delta_4 h^{\nu \sigma } \delta_3 h_{\nu\sigma}+20 \delta_2 h^{\mu \rho } \delta_4 h^{\nu \sigma } \delta_3 h_{\nu\sigma}+4 d \delta_2 h^{\mu }_{\sigma } \delta_4 h^{\nu \rho } \delta_3 h^{\sigma }_{\nu }\nn\\
&-32 \delta_2 h^{\mu }_{\sigma } \delta_3 h^{\rho }_{\nu } \delta_4 h^{\nu \sigma }+16 \delta_2 h^{\mu }_{\sigma } \delta_4 h^{\nu \rho } \delta_3 h^{\sigma }_{\nu }\Big)\Big]\,.
\ee
The corresponding four-point contact diagram is then evaluated by
\be
\mathcal{M}_{4,{\rm grav}}^{\rm ct}= \int D^{d+1}Y \big(V_{1234,h}(Y)+{\rm perm}\big)\,.\label{eq: 4-pt ct}
\ee

\section{On the graviton bulk-to-boundary propagator}
\label{app: ddh rep and id}

In this appendix, we complete the differential representation for bulk-to-boundary propagators \eqref{eq: ddh rep}. We provide detailed identities for contracting bulk-to-boundary propagators when deriving the differential representation.

The complete version of \eqref{eq: ddh rep} is
\be
&\delta_i h_{\mu\nu} = \mathcal{E}_{i,\mu}\mathcal{E}_{i,\nu} \delta_i \phi_{d}\,,\nn\\
& \nabla_{\mu}\delta_i h_{\nu\rho} = \mathcal{E}_{i,\nu}\mathcal{E}_{i,\rho} \mathcal{P}_{i\mu}\delta_i \phi_{d-1}-\big(Y_{\rho } \mathcal{E}_{i \mu } \mathcal{E}_{i \nu }\delta_i\phi _d +Y_{\nu } \mathcal{E}_{i \mu } \mathcal{E}_{i \rho }\delta_i\phi_d\big)\,,\nn\\
& \nabla_{\mu}\nabla_{\nu}\delta_i h_{\rho\sigma}=\mathcal{E}_{i,\rho}\mathcal{E}_{i,\sigma} \mathcal{P}_{i\mu}\mathcal{P}_{i\nu} \delta_i \phi_{d-2} -Y_{\sigma } \mathcal{E}_{i \nu } \mathcal{E}_{i \rho }\mathcal{P}_{i \mu } \delta_i\phi_{d-1}-Y_{\mu } \big(\mathcal{E}_{i \rho } \mathcal{E}_{i \sigma } \mathcal{P}_{i \nu } \delta_i\phi_{d-1}+Y_{\sigma } \mathcal{E}_{i \nu } \mathcal{E}_{i \rho }\delta_i\phi_d\big)\nn\\
&-Y_{\rho } \big(\mathcal{E}_{i \nu } \mathcal{E}_{i \sigma }\mathcal{P}_{i \mu } \delta_i\phi_{d-1}+Y_{\mu } \mathcal{E}_{i \nu } \mathcal{E}_{i \sigma }\delta_i\phi_d\big)-\delta _{\mu\rho }\mathcal{E}_{i \nu } \mathcal{E}_{i \sigma }\delta_i\phi_d -\delta _{\sigma\mu }\mathcal{E}_{i \nu } \mathcal{E}_{i \rho }\delta_i\phi_d \,.\label{eq: compete ddh rep}
\ee

To arrive at differential representation exhibiting flat-space structure, we should prove identities with the spirit of transverse-traceless and on-shell conditions in flat-space. We find
\be
 \mathcal{E}_i\cdot \mathcal{E}_i \,\mathcal{P}_{i\mu}\mathcal{P}_{i\nu}\delta_i\phi_{d-2}= \mathcal{E}_i\cdot \mathcal{E}_i\delta_i\phi_{d}=\mathcal{P}_i\cdot \mathcal{P}_i \delta_i\phi_{d-2}= \mathcal{E}_i\cdot \mathcal{P}_i \,\mathcal{P}_{i\mu}\delta_i\phi_{d-2}= \mathcal{E}_i\cdot \mathcal{P}_i\delta_i\phi_{d-1}=0\,,
\ee
For those terms $\mathcal{O}(Y), \mathcal{O}(Y,g)$, we prove a set of identities that can help us get rid of $Y$ in the final representation so that the differential representation is well-defined even from perspective pure CFT

\be
& Y\cdot\mathcal{P}_i\delta_i\phi_{d-1}=(Y\cdot\mathcal{P}_i)^2\delta_i\phi_{d-2}=Y\cdot\mathcal{E}_i\delta_i\phi_d=(Y\cdot\mathcal{E}_i)^2\delta_i\phi_d=(Y\cdot\mathcal{E}_i)^2\,\mathcal{P}_{i,\mu}\delta_i\phi_{d-1}=0\,,\nn\\
& Y\cdot\mathcal{P}_i\,\mathcal{P}_{i,\mu}\delta_i\phi_{d-2}=-\mathcal{P}_{i,\mu}\delta_i\phi_{d-1}\,,\quad (Y\cdot\mathcal{E}_i)^2\,\mathcal{P}_{i,\mu}\mathcal{P}_{i,\nu}\delta_i\phi_{d-2}=2\mathcal{E}_{i,\mu}\mathcal{E}_{i,\nu}\delta_i\phi_{d}\,,\nn\\
& Y\cdot\mathcal{E}_i\,\mathcal{P}_{i,\mu}\delta_i\phi_{d-1}=-\mathcal{E}_{i,\mu}\delta_i\phi_{d}\,,\quad Y\cdot\mathcal{E}_i\,\mathcal{P}_{i,\mu}\mathcal{P}_{i,\nu}\delta_i\phi_{d-2}=-\mathcal{E}_{i,(\mu}\mathcal{P}_{i,\nu)}\delta_i\phi_{d-1}\,.
\ee

\section{On the graviton bulk-to-bulk propagator}
\label{app: bulk-to-bulk}

In this appendix, we show in detail how we derive \eqref{eq: D12 graviton} by following the lines of \cite{Herderschee:2022ntr, YueZhou:future}. For simplicity, we take the de-Donder gauge for the propagating graviton. It is useful to decompose the graviton into the traceless part and the trace part
\be
h_{\mu\nu} = \tilde{h}_{\mu\nu} +\fft{1}{d+1} h g_{\mu\nu}\,,\quad \tilde{h}\equiv 0\,.
\ee
The basic idea is to treat $\tilde{h}_{\mu\nu}$ and $h$ independently.They give rise to three different bulk-to-bulk propagators by the Wick contractions that satisfy different equations. To see this, we consider the equation of motion of graviton in the de-Donder gauge
\be
(\nabla_{Y_1}^2+2) \left\langle h_{\mu\nu}(Y_1)h_{\rho\sigma}(Y_2)\right\rangle-
2g_{\mu\nu} \left\langle h(Y_1)h_{\rho\sigma}(Y_2)\right\rangle=\fft{1}{2}\big(g_{\mu\rho}g_{\nu\sigma}+g_{\mu\sigma}g_{\nu\rho}-\fft{2g_{\mu\nu}g_{\rho\sigma}}{d-1}\big)\delta(Y_1-Y_2)\,.
\ee
It is easy to find that upon the trace decomposition \eqref{eq: trace decomp} we have
\begin{align}
& (\nabla_{Y_1}^2+2) \left\langle \tilde{h}_{\mu\nu}(Y_1)\tilde{h}_{\rho\sigma}(Y_2)\right\rangle=\fft{1}{2}\big(g_{\mu\rho}g_{\nu\sigma}+g_{\mu\sigma}g_{\nu\rho}-\fft{2g_{\mu\nu}g_{\rho\sigma}}{d+1}\big)\delta(Y_1-Y_2)\,,\nn\\
& (\nabla_{Y_1}^2-2d)\left\langle h(Y_1)h(Y_2)\right\rangle = \fft{2(d+1)}{d-1}\delta(Y_1-Y_2)\,,\nn\\
& (\nabla_{Y_1}^2-2d)\left\langle h(Y_1)\tilde{h}_{\mu\nu}(Y_2)\right\rangle = (\nabla_{Y_1}^2+2)\left\langle h_{\mu\nu}(Y_1)h(Y_2)\right\rangle =0\,.\label{eq: prop eqs}
\end{align}
Besides, any vertex functions coupled to the stress-tensor are now naturally decomposed into traceless and trace part
\be
V_{\mu\nu}h^{\mu\nu} = \tilde{V}_{\mu\nu} \tilde{h}^{\mu\nu} + \fft{1}{d+1} {\rm Tr}\,V\, h\,.
\ee

With all these ingredients in mind, we can rewrite the graviton exchange amplitudes {eq: graviton exchange diagram} as
\begin{align}
 \mathcal{M}^{(s)}_{4{\rm ex},{\rm grav}}=& 16\times 8\pi G\int D^{d+2}Y_1D^{d+2}Y_2\, \Big(\tilde{V}_{h,12}^{\mu\nu}\left\langle \tilde{h}_{\mu\nu}(Y_1)\tilde{h}_{\rho\sigma}(Y_2)\right\rangle \tilde{V}_{h,34}^{\rho\sigma}+ \fft{1}{(d+1)^2}{\rm Tr}\,V_{h,12} \left\langle h(Y_1)h(Y_2)\right\rangle {\rm Tr}\, V_{h,34}\nn\\
&+\fft{1}{d+1}{\rm Tr}\,V_{h,12}\left\langle h(Y_1)\tilde{h}_{\rho\sigma}(Y_2)\right\rangle \tilde{V}_{34,h}^{\rho\sigma}+\fft{1}{d+1}\tilde{V}_{h,12}^{\mu\nu}\left\langle \tilde{h}_{\mu\nu}(Y_1)h(Y_2)\right\rangle {\rm Tr}\,V_{h,34} \Big)\,.
\end{align}
Using the AdS embedding formalism for the bulk-to-boundary propagator \eqref{eq: bulk-to-boundary graviton}, we can easily prove the following identities
\be
\mathcal{C}_{12} \tilde{V}_{h,12}^{\mu\nu}= \big(\nabla_{Y_1}^2+2(d+1)\big)\tilde{V}_{h,12}^{\mu\nu}\,,\quad \mathcal{C}_{12}{\rm Tr}\,V_{h,12} = \nabla_{Y_1}^2 {\rm Tr}\,V_{h,12}\,.
\ee
Therefore, we find 
\begin{align}
& \mathcal{D}^{d,2}_{12} \mathcal{M}_{4{\rm ex},{\rm grav}} = \int D^{d+2}Y_1 D^{d+2}Y_2\times\nn\\
& \Big(\tilde{V}^{\mu\nu}_{h,12} (\nabla_{Y_1}^2+2)\left\langle \tilde{h}_{\mu\nu}(Y_1)\tilde{h}_{\rho\sigma}(Y_2)\right\rangle \tilde{V}_{h,34}^{\rho\sigma} + \fft{1}{(d+1)^2} {\rm Tr}\,V_{h,12} (\nabla_{Y_1}^2-2d) \left\langle h(Y_1)h(Y_2)\right\rangle {\rm Tr}\, V_{h,34} \nn\\
& +\fft{1}{d+1}{\rm Tr}\,V_{h,12}(\nabla_{Y_1}^2-2d)\left\langle h(Y_1)\tilde{h}_{\rho\sigma}(Y_2)\right\rangle \tilde{V}_{h,34}^{\rho\sigma}+\fft{1}{d+1}\tilde{V}_{h,12}^{\mu\nu}(\nabla_{Y_1}^2+2)\left\langle \tilde{h}_{\mu\nu}(Y_1)h(Y_2)\right\rangle {\rm Tr}\,V_{h,34} \Big)\,.\label{eq: trace decomp}
\end{align}
Plugging \eqref{eq: prop eqs} into \eqref{eq: trace decomp}, we prove \eqref{eq: D12 graviton}.

\end{appendix}

\end{document}